\documentclass[11pt]{article}
\usepackage{graphicx}
\usepackage{natbib} 
\usepackage{url} 
\usepackage{float}
\usepackage{amsfonts}
\usepackage{amsmath}
\usepackage{booktabs}
\newtheorem{theorem}{Theorem}
\RequirePackage[ruled,vlined,linesnumbered]{algorithm2e}
\usepackage[colorlinks=true,linkcolor=blue, citecolor=blue]{hyperref}
\usepackage{comment}
\usepackage{amssymb}
\usepackage[color]{changebar}
\cbcolor{red}
\newcommand{\blind}{0}

\newcommand{\ra}[1]{\renewcommand{\arraystretch}{#1}}

\usepackage{booktabs,natbib,tabularx,ragged2e,array}
\newcolumntype{Y}{>{\RaggedRight\arraybackslash\setlength{\parskip}{0pt}}X} 
\def\bSig\mathbf{\Sigma}

\newcommand{\ee}{\mathbb{E}}
\newcommand{\rr}{\mathcal{R}}
\newcommand{\qed}{\tag*{$\square$}}

\newcommand{\meansd}[2]{#1\,\ensuremath{\pm}\,#2}

\begin{document}

\def\spacingset#1{\renewcommand{\baselinestretch}%
{#1}\small\normalsize} \spacingset{1}

\if0\blind
{
\begin{titlepage}
\thispagestyle{empty}

\begin{center}
  \parbox{\textwidth}{\raggedright
    {\large \textbf{Sensitivity Analysis when Generalizing Causal Effects from Multiple Studies to a Target Population: Motivation from the ECHO Program
}}
  }
\end{center}



\begin{center}
  \parbox{\textwidth}{\raggedright
    Bolun Liu$^{1}$,
    Trang Quynh Nguyen$^{2}$,
    Elizabeth A. Stuart$^{1,*}$,
    Bryan Lau$^{3}$,
    Amii M. Kress$^{3}$,
    Michael R. Elliott$^{4}$,
    Kyle R. Busse$^{5}$,
    Ellen C. Caniglia$^{5}$,
    Yajnaseni Chakraborti$^{5}$,
    Amy J. Elliott$^{6}$,
    James E. Gern$^{7}$,
    Alison E. Hipwell$^{8}$,
    Catherine J. Karr$^{9}$,
    Kaja Z. LeWinn$^{10}$,
    Li Luo$^{11}$,
    Hans\mbox{-}Georg Müller$^{12}$,
    Sunni L. Mumford$^{5}$,
    Ruby H. N. Nguyen$^{13}$,
    Emily Oken$^{14}$,
    Janet L. Peacock$^{15}$,
    Enrique F. Schisterman$^{5}$,
    Arjun Sondhi$^{16}$,
    Rosalind J. Wright$^{17}$,
    Yidong Zhou$^{12}$,
    Elizabeth L. Ogburn$^{1}$
  }
\end{center}

\begin{center}
  \parbox{\textwidth}{\raggedright\scriptsize
    $^{1}$Department of Biostatistics, Johns Hopkins Bloomberg School of Public Health, Baltimore, Maryland, USA\\
    $^{2}$Department of Mental Health, Johns Hopkins Bloomberg School of Public Health, Baltimore, Maryland, USA\\
    $^{3}$Department of Epidemiology, Johns Hopkins Bloomberg School of Public Health, Baltimore, Maryland, USA\\
    $^{4}$Department of Biostatistics, University of Michigan School of Public Health, Ann Arbor, Michigan, USA\\
    $^{5}$Department of Biostatistics, Epidemiology, and Informatics, University of Pennsylvania, Philadelphia, Pennsylvania, USA\\
    $^{6}$Avera Research Institute, Avera Health, Sioux Falls, South Dakota, USA\\
    $^{7}$Department of Pediatrics, School of Medicine and Public Health, University of Wisconsin–Madison, Madison, Wisconsin, USA\\
    $^{8}$Department of Psychiatry, Psychology, and Clinical and Translational Science, University of Pittsburgh, Pittsburgh, Pennsylvania, USA\\
    $^{9}$Department of Pediatrics and Department of Environmental and Occupational Health Sciences, University of Washington, Seattle, Washington, USA\\
    $^{10}$Department of Psychiatry and Behavioral Sciences, University of California, San Francisco, San Francisco, California, USA\\
    $^{11}$Department of Internal Medicine and Comprehensive Cancer Center, University of New Mexico, Albuquerque, New Mexico, USA\\
    $^{12}$Department of Statistics, University of California, Davis, Davis, California, USA\\
    $^{13}$Division of Epidemiology and Community Health, University of Minnesota, Minneapolis, Minnesota, USA\\
    $^{14}$Department of Population Medicine, Harvard Pilgrim Health Care Institute, Boston, Massachusetts, USA\\
    $^{15}$Department of Epidemiology, Dartmouth College, Hanover, New Hampshire, USA\\
    $^{16}$Feinstein Institutes for Medical Research, Northwell Health, New York, New York, USA\\
    $^{17}$Department of Public Health, Icahn School of Medicine at Mount Sinai, New York, New York, USA\\[0.7em]
    $^{*}$Corresponding author: Elizabeth A. Stuart, 624 N Broadway, Baltimore, MD 21205, USA; \texttt{estuart@jhsph.edu}.
  }
\end{center}

\end{titlepage}
} \fi

\if1\blind
{
\begin{titlepage}
\thispagestyle{empty}
\begin{center}
  \parbox{0.92\textwidth}{\raggedright
    {\LARGE\bfseries Sensitivity Analysis when Generalizing Causal Effects from Multiple Studies to a Target Population}
  }
\end{center}
\end{titlepage}
} \fi

\bigskip
\begin{abstract}
Unobserved effect modifiers can induce bias when generalizing causal effect estimates to target populations. In this work, we extend a sensitivity analysis framework assessing the robustness of study results to unobserved effect modification that adapts to various generalizability scenarios, including multiple (conditionally) randomized trials, observational studies, or combinations thereof. This framework is interpretable and does not rely on distributional or functional assumptions about unknown parameters. We demonstrate how to leverage the multi-study setting to detect violation of the generalizability assumption
through hypothesis testing, showing with simulations that the proposed test achieves high power under real-world sample sizes. Finally, we apply our sensitivity analysis framework to analyze the generalized effect estimate of secondhand smoke exposure on birth weight using cohort sites from the Environmental influences on Child Health Outcomes (ECHO) study.
\end{abstract}

\noindent%
\vfill

\newpage
\spacingset{1.5}

\section{Introduction}
\label{s:intro}
External validity of biomedical studies -- the extent to which causal effects estimated in study samples can be generalized or transported to specific target populations -- is essential for informing clinical decisions and public health policies, particularly when directly estimating the causal effects in the target populations are infeasible, prohibitively expensive, or ethically challenging. Methods for extending inference from a single study are relatively well-developed, typically involving adjustments for the distributional differences in pretreatment covariates between the study samples and the target population \citep{lizweightest, issasingle, karatransport}. These methods often require adjusting for a complete set of effect modifiers, i.e., covariates that interact with the treatment assignment 
and contribute to treatment effect heterogeneity. However, such a requirement is seldom met in practice, which can lead to significant bias \citep{lizassess, echoshs}. This necessitates the development of sensitivity analyses to quantify the impact of unobserved effect modifiers on the generalized effect estimates. 
 
Although various sensitivity analysis methods have been developed (see Section~\ref{subs: Literature}), relatively few have explored how to adapt these approaches for multi-study settings, where data from multiple (and potentially conditionally randomized) study populations are combined for generalization. Such multi-study designs are increasingly common in public health research, enabling investigators to synthesize evidence across diverse populations and settings, thus broadening the applicability of causal findings beyond any single study \citep{issatowards}. A prime example is the Environmental influences on Child Health Outcomes (ECHO) Program, which has consolidated data from more than 60,000 children from 69 cohorts in 49 states, the District of Columbia, and Puerto Rico, facilitating an in-depth examination of how exposures from pregnancy through the early years of development affect child health trajectories \citep{echo}. To capitalize on the efficiencies of ongoing cohort studies, the National Institutes of Health (NIH) initially funded sites already participating in pregnancy and pediatric cohorts, bringing them together to form ECHO as a multi-study consortium. Beyond examining how physical, chemical, biological, and social determinants shape child health within ECHO, there is growing interest in whether these findings can be generalized to broader populations, such as children across the United States. Given the complexities of merging and generalizing evidence across multiple cohort studies, particularly due to conditional randomization and variation in treatment ratios, sensitivity analyses developed for single-trial generalization settings are no longer directly applicable. Thus, there is a clear need for sensitivity analyses specifically designed for these multi-study contexts such as ECHO, to address challenges posed by potential unobserved effect modification.
  

In this paper, we propose a sensitivity analysis framework for assessing the implications of potential unobserved effect modifiers that are missing in both the source and target samples when generalizing causal effects from multiple studies to a target population. By adapting the non-parametric sensitivity framework of \cite{huang} to a multi-study weighted estimator modified from \cite{casmeta}, our method can accommodate a variety of generalizability scenarios --- including multiple (conditionally) randomized trials, observational studies, or combinations thereof. This extended framework retains the advantages of \cite{huang}, which does not rely on distributional or functional assumptions about unknown parameters, and inherits graphical and formal benchmarking tools to enhance the interpretability of sensitivity analyses. Furthermore, we give suggestions on how to leverage the multi-study setting to detect unobserved effect modification through hypothesis testing, and show with simulation studies that the proposed test achieves high power under real-world scenarios. 

The paper is structured as follows. First, we review relevant literature in the remainder of this section. Next, we introduce the notation, set-up, and the multi-study weighted estimator in Section \ref{s:setup}. In Section \ref{s:sens}, we present the key theorem that extends sensitivity analysis to the multi-study setting, and elaborate the estimation details of sensitivity and benchmarking parameters. We then discuss hypothesis testing for unobserved effect modification along with simulation results in Section \ref{s:hypo}. Lastly, in Section \ref{s:echo}, we apply our sensitivity analysis framework to generalizing the effect of secondhand smoke (SHS) exposure on birth weight using cohort sites from the ECHO consortium. 

\subsection{Methodological Background}
\label{subs: Literature}

Generalizing causal effects from a study sample to a target population typically requires adjusting estimates to reflect differences in the distribution of effect modifiers between the study sample and the target population.
To address this issue, \cite{colestuart} first introduced a model-based method to adjust for distributional differences in covariates between the study and the target population, which led to the development of various generalizability methods. \cite{tiptonreview} and \cite{generalizabilityreview} provide comprehensive reviews of recent advances in generalizability methods. When multiple studies are available, \cite{issatowards} proposed an inverse probability weighted estimator that combines evidence across studies for generalization. Building on this, \cite{casmeta} further developed doubly robust, causally interpretable meta-analysis frameworks under a variety of identifying assumptions.

To address the challenge of unobserved effect modifiers, \cite{trangfirstses} and \cite{trangsecondsen} pioneered sensitivity analyses for unobserved effect modifications in the generalizability setting. Their work focuses on outcome models with additive or multiplicative effects and assumes independence between the unobserved and observed effect modifiers. Extending this line of research, \cite{Colnet} developed sensitivity analysis for semi-parametric models, assuming that the unobserved modifiers are independent and normally distributed. \cite{nie} proposed a non-parametric sensitivity framework for a Horvitz-Thompson type weighted estimator, which uses the ratio of generalization weights (formally defined in equation (\ref{weight})) as sensitivity parameters, without imposing independent or distributional assumptions on the unobserved modifiers. \cite{huang} also examined the same estimator, but analyzed the generalization weight differences instead, and further provided tools to benchmark the impact of unobserved versus observed modifiers on generalized effect estimates. Using a different approach, \cite{issabias} considered bias functions to assess any violation of the generalizability assumption (Assumption A5 in Section \ref{s:setup}), which subsumes the case where there are unobserved effect modifiers. Building on this, \cite{duong} extends the bias function-based sensitivity analysis to the multi-study setting. However, similar to \cite{issabias}, this method relies on the correct specification of the bias functions, which may be challenging in practice.

In summary, existing sensitivity frameworks either primarily focus on single-trial generalization scenarios, which are unsuitable for multi-study generalization situations, especially in the presence of conditional randomization or varying treatment ratios across studies, or rely on additional independence or parametric assumptions for unknown parameters. This gap underscores the need for our proposed nonparametric, multi-study sensitivity analysis framework.

\section{Set-up \& Notation}
\label{s:setup}

Suppose we observe a collection of trials or observational studies indexed by \( S = 1, \cdots, m \), with sample sizes \( n_1, \cdots, n_m \). For each participant in the collection of studies,
we have information on study membership \( S \), binary treatment assignment \( A \in \{0, 1\} \), outcome \( Y \). We assume that \( X \) includes treatment effect modifiers \( V \) (i.e. $V \subseteq X$) and covariates contributing to conditional ignorability, i.e., covariates defining conditional strata in conditionally randomized trials, or confounders in observational studies. Additionally, we assume that we observe the full population or a simple random sample from a target population of interest, indexed by \( S = 0 \), with sample size \( n_0 \) and only covariates $X$ observed, but not treatment or outcome. Thus, the total number of units in the studies and the target population sample is \( n \equiv \sum_{s = 0}^m n_s \). Let $R$ denote the indicator variable for whether the unit is in the collection of studies (vs. the target population sample), i.e., $R = 1$ if and only if $S \neq 0$.


Following \cite{casmeta}, we assume the tuples \((S_i, X_i, R_i \times A_i, R_i \times Y_i)\), \( i = 1, \cdots, n \), are independent and identically distributed random samples.
We further assume that participants are exclusively assigned to one study or target population sample, receive identical (or harmonized, \cite{Volk2021-bk}) versions of the treatment and control conditions, and have their pretreatment covariates and outcomes measured consistently across all studies and target population samples. For each participant $i$, let $Y^a_i$ denote the potential outcomes under treatment $a \in \{0, 1\}$ \citep{potentialoutcome}.
Let $\tau_i \equiv Y_i^1 - Y_i^0$ denote the individual treatment effect, and the estimand of interest is the target population average treatment effect (PATE)
\[\tau \equiv \mathbb{E}(\tau_i \mid R_i = 0)\;.\]

We consider a generalization scenario where we only require that the union of the covariate supports across all studies covers that of the target population, which enables generalization even when individual studies do not fully overlap with the target. For example, consider two studies conducted separately in youth and adults. While neither study can, on its own, generalize to a target population including both age groups due to lack of overlap, their findings can be combined to yield valid generalizations. Consequently, it allows the inclusion of as many studies as possible in the analysis, thereby maximizing the use of available information \citep{casmeta}. It also facilitates generalization from conditionally randomized trials, as we may naturally view a conditionally randomized trial as a collection of non-overlapping trials indexed by the variables that define the conditional strata used in the randomization. 

We require the following assumptions for identification of PATE. 
\begin{itemize}
    \item[A1] Consistency: $Y = AY^1 + (1 - A)Y^0 \;.$
    \item[A2] Positivity of Participation: If $p(X = x \mid R = 0) \neq 0$, then $P(X = x \mid R = 1) > 0 \;.$
    \item[A3] Within Study Ignorable Treatment Assignment: For each $s = 1, \dots, m$ and $a=0,1$, $Y^a \perp A \mid (X, S = s) \;.$
    \item[A4] Within Study Positivity of Treatment Assignment: For each $s = 1, \cdots, m$, if $p(X = x \mid S = s) \neq 0$, then $P(A = 1 \mid X = x, S = s) > 0 \;.$
    \item[A5] Exchangeability of effect measure over $S$: For each $s = 1, \cdots, m$, $E(Y^1 - Y^0\mid V, S = s) = E(Y^1 - Y^0 \mid V, R = 0) \;.$
\end{itemize}

Specifically, underlying A1, we assume the Stable Unit Treatment Value Assumption (SUTVA) \citep{sutva1, sutva2}, i.e., there is a single version of each treatment, no interference or spillover effect between individuals within or across studies, and no direct effect of participation in any study $(R_i = 1)$ or participation in some specific study $(S_i = s)$ on the outcomes; i.e., the potential outcomes are the same for a given individual regardless of the study or sample they might be in \citep{consist1, issasingle, casmeta}. A2 addresses the general covariate overlap assumption, requiring that at least one study represents each part of the target population. A3 typically arises from (conditional) randomization (and in that case it is satisfied by the randomization), but also applies to a collection of observational studies; in that case it will be satisfied if the covariates \(X\) are sufficient to adjust for baseline confounding. A4 assumes positivity of treatment assignment within each study or within the conditional strata. A5, often referred to as the generalizability assumption, states that conditioning on $V$ adjusts for all effect modifiers that distribute differently in the studies and target population, thus allowing the conditional average treatment effects to be generalized from each study to the target population.


When assumptions A1 through A5 hold, the following theorem establishes identification of the PATE, which generalizes upon the results from \cite{casmeta}. 
\begin{theorem}
    \label{thm1}
    Under assumptions A1--A5, the target population average treatment effect $\tau = \mathbb{E}(Y^1_i - Y^0_i \mid R_i = 0)$ is identified as    
    \begin{equation}
        \begin{aligned}
            \mathbb{E}(Y_i^1 - Y_i^0 & \mid R_i = 0) = \frac{1}{P(R_i=0)} \mathbb{E}\left[\frac{P(R_i = 0 \mid V_i)}{P(R_i = 1 \mid V_i)} \right. \\
    &\left. \left(\frac{I(R_i=1, A_i=1)Y_i}{P(A_i=1 \mid X_i, S_i, R_i=1)} - \frac{I(R_i=1, A_i=0)Y_i}{P(A_i=0 \mid X_i, S_i, R_i=1)} \right)\right] \;.
        \end{aligned}
    \end{equation}
    
\end{theorem}
Proof of Theorem \ref{thm1} is provided in Appendix \hyperref[app: thm1]{A}. Theorem \ref{thm1} establishes identification of the PATE under weak overlap and generalizability assumptions, and further shows that generalization weights $w_i$ need only adjust for the complete set of effect modifiers $V$ rather than the full covariate set $X$. This refinement, consistent with \cite{karaeffi}, can yield more efficient estimators when effect modifiers constitute only a small subset of the covariates.


From Theorem \ref{thm1}, a Horvitz–Thompson type weighted estimator for the PATE follows naturally, which takes the form: 
    \begin{equation}
        \label{eq: estimator}
        \widehat{\tau}_W \equiv \frac{1}{N_1}\sum_{i \in \mathcal{R}} w_i \lambda_i \gamma_i A_i Y_i - \frac{1}{N_0}\sum_{i \in \mathcal{R}} w_i \lambda_i \gamma_i (1 - A_i) Y_i \;,
    \end{equation}
    where $N_a$ denotes the total number of individuals receiving treatment $A = a$ in the collection of studies, the subscript $\mathcal{R}$ indicates the sum is over all samples in the collection of studies, and $w_i$, $\lambda_i$, $\gamma_i$ represents the large-sample limits of the weights
    \begin{equation}
    \label{weight}
        \begin{aligned}
        w_i &\equiv \frac{P(R_i = 1)}{P(R_i = 0)} \frac{P(R_i=0 \mid V_i)}{P(R_i=1 \mid V_i)}\;, 
        \quad \lambda_i = \frac{P(A_i \mid R_i = 1)}{P(A_i \mid S_i, R_i = 1)}\;, \\
         &\quad \quad \quad \quad \quad \text{and} \quad \gamma_i \equiv \frac{P(A_i \mid S_i, R_i = 1)}{P(A_i \mid X_i, S_i, R_i=1)} \;.
\end{aligned}
    \end{equation}

We denote $w_i$ as the \textit{generalization weight}, which ensures generalizability of effect estimates by adjusting for the distributional difference in the effect modifiers between the collection of studies and target population. We denote $\lambda_i$ as the \textit{combination weight}, which adjusts for the different treatment ratios across the studies. We denote $\gamma_i$ as the \textit{de-confounding weight}, which controls for confounding within each study by balancing the covariate distributions across the treatment and control groups. Informally, the proposed estimator first emulates each single study (or conditional strata of conditional randomized trial) as a randomized (or essentially randomized, assuming A2) trial through the de-confounding weights $\gamma_i$, then aligns them with the same treatment ratio through the combination weight $\lambda_i$, allowing them to be pooled as a unified trial. It then generalizes from the unified trial to the target population through the generalization weights $w_i$, akin to the approach used in a single-trial generalization scenario.  Consequently, in situations where there is either a single randomized trial or multiple trials with the same treatment ratio, we observe that $\lambda_i = \gamma_i = 1$, due to the same treatment ratios and randomization, respectively. This simplifies $\widehat{\tau}_W$ to the weighted estimator that generalizes from a single trial to the target population described in \cite{lizweightest} and \cite{huang}. Thus, the proposed estimator in Theorem \ref{thm1} can be viewed as a generalized version of the single-trial generalizability estimator and can be used for generalizing from a single trial, a single observational study, multiple trials, multiple observational studies, or any combination thereof.

\section{Sensitivity Analysis for Multiple-Study Weighted Estimators}
\label{s:sens}
We now consider the sensitivity of the proposed multi-study weighted estimator when an (or a set of) effect modifier is omitted in the estimation of the weights $w_i, \lambda_i$, and $\gamma_i$. Suppose the complete set of effect modifiers is $\{V, U\}$. However, for some reason, $U$ is missing from the data collection process, leaving only $V$ available for estimation of the weights. To facilitate the sensitivity analysis, we now require that $U$ not act as a confounder. This assumption automatically holds when generalizing from a collection of randomized trials, where randomization eliminates confounding even if $U$ is unobserved. However, when generalizing from observational studies, it requires that A2 holds conditioning only on $X$.

If $U$ were observed in the collection of studies and the target population sample, the PATE could be estimated using the \textit{ideal} weights defined with a complete set of effect modifiers $\{V, U\}$, written as
\begin{equation}
    \label{idealw}
        \begin{aligned}
        w_i^* &\equiv \frac{P(R_i = 1)}{P(R_i = 0)} \frac{P(R_i=0 \mid V_i, U_i)}{P(R_i=1 \mid V_i, U_i)}\;, 
        \quad \lambda_i^* \equiv \frac{P(A_i \mid R_i = 1)}{P(A_i \mid S_i, R_i = 1)}\;, \\
         &\quad \quad \quad \quad \quad \text{and} \quad \gamma_i^* \equiv \frac{P(A_i \mid S_i, R_i = 1)}{P(A_i \mid X_i, U_i, S_i, R_i=1)} \;.
    \end{aligned}
\end{equation}
However, since $U$ is not observed, the PATE is estimated using the \textit{misspecified} weights with the observed effect modifier $V$, written as 
    \begin{equation}
        \label{misw}
        \begin{aligned}
        w_i &\equiv \frac{P(R_i = 1)}{P(R_i = 0)} \frac{P(R_i=0 \mid V_i)}{P(R_i=1 \mid V_i)}\;, 
        \quad \lambda_i \equiv \frac{P(A_i \mid R_i = 1)}{P(A_i \mid S_i, R_i = 1)}\;, \\
         &\quad \quad \quad \quad \quad \text{and} \quad \gamma_i \equiv \frac{P(A_i \mid S_i, R_i = 1)}{P(A_i \mid X_i, S_i, R_i=1)} \;.
\end{aligned}
\end{equation}
Consequently, when $w_i^*,w_i^*,\lambda_i^*,\lambda_i,\gamma_i^*,\gamma_i$ represent the large sample limits of the inverse propensity score weights, the disparity between the ideal and misspecified weights due to the unobserved $U$ becomes the primary source of bias, and our sensitivity analysis will focus specifically on the bias arising from this weight misspecification.

We now present the following theorem, which states that the bias from omitting an effect modifier in the estimation of the weights can be expressed as a function of the generalization weights $w_i^*, w_i$ and individual treatment effects $\tau_i$.

\begin{theorem}
    \label{thm2}
    Let \( w_i^* \) and \( w_i \) be defined as in equations (\ref{idealw}) and (\ref{misw}), respectively, and let \( \tau_i \) denote the individual treatment effect. The expectations of the misspecified and ideal (unbiased) weighted estimators can be expressed as functions of only the generalization weights and individual treatment effects:
    \begin{equation}
        \ee(\widehat{\tau}_W) = \ee_\rr(w_i \tau_i) \quad \text{and} \quad \ee(\widehat{\tau}_W^*) = \ee_\rr(w_i^* \tau_i) = \tau \;, 
    \end{equation}
    where the subscript $\mathcal{R}$ indicates the expectation is taken over the study samples, i.e., $\mathbb{E}_\mathcal{R}(\cdot) = \mathbb{E}(\cdot \mid R_i = 1)$, $\operatorname{cov}_\mathcal{R}(\cdot) = \operatorname{cov}(\cdot \mid R_i = 1)$, etc.. Further denote the linear difference between the misspecified and ideal weights as $\varepsilon_i \equiv w_i - w^*_i$. The asymptotic bias of the multi-study weighted estimator from using $w_i$ instead of $w_i^*$ can be written as
    \begin{equation}
        \label{biasdecompose}
        \operatorname{Bias}\left(\widehat{\tau}_W\right) = \operatorname{cov}_\rr(\varepsilon_i, \tau_i) \;.
    \end{equation}
\end{theorem}
Proof of Theorem \ref{thm2} is provided in Appendix \hyperref[app: thm2]{B}.

Theorem \ref{thm2} states that the bias of the multi-study weighted estimator arises exclusively from misspecification of the generalization weights and remains unaffected by the combination or de-confounding weights. Intuitively, the unobserved effect modifier $U$ does not alter the combination weight because $\lambda_i$ only depends on the treatment ratios within each study, so that $\lambda_i^* = \lambda_i$. Moreover, since the observed covariates $X$ are assumed to be sufficient for adjusting for confounding within each study (i.e., assumption A3), the de-confounding weights $\gamma_i^*$ and $\gamma_i$ are conceptually equivalent regardless of whether $U$ is observed. Consequently, any misspecification of the generalization weight $w_i$ becomes the only source of bias. As a result, the bias decomposition of the multi-study weighted estimator parallels that of the single-trial weighted estimator described in \cite{huang}, and the single-trial sensitivity framework can be directly generalized to the multi-study scenario.

\subsection{Single-Trial Sensitivity Framework in Huang (2024)}

We now introduce the sensitivity framework in \cite{huang}, developed for a setting with a single trial being generalized to a target population. Building on the decomposition in equation (\ref{biasdecompose}), \cite{huang} further writes the bias of the weighted estimator as a function of three sensitivity parameters. Specifically, denote 
\begin{equation}
\label{senparam}
    R_\varepsilon^2 \equiv \operatorname{var}_\rr (\varepsilon_i) / \operatorname{var}_\rr (w_i^*),\quad \rho_{\varepsilon, \tau} \equiv \operatorname{cor}_\rr(\varepsilon_i, \tau_i), \quad \text{and} \quad \sigma_\tau^2 \equiv \operatorname{var}_\rr(\tau_i) \;,
\end{equation}
then the bias in (\ref{biasdecompose}) can be rewritten as 
\begin{equation}
\label{biasformula}
    \operatorname{Bias}\left(\widehat{\tau}_W\right)=\left\{\begin{array}{cc}
    \rho_{\varepsilon, \tau} \sqrt{R_{\varepsilon}^2 / \left(1-R_{\varepsilon}^2 \right) \cdot \operatorname{var}_{\mathcal{S}}\left(w_i\right) \cdot \sigma_\tau^2} & \text { if } R_{\varepsilon}^2<1 \\
    \rho_{\varepsilon, \tau} \sqrt{\operatorname{var}_{\mathcal{S}}\left(w_i^*\right) \cdot \sigma_\tau^2} & \text { if } R_{\varepsilon}^2=1 \;.
    \end{array}\right.
\end{equation}

The $R^2_\varepsilon$ term serves as a measure of the residual imbalance in the unobserved effect modifier $U$ after controlling for the observed effect modifier $V$, which ranges from 0 to 1, as $\operatorname{var}_\rr(w_i^*) = \operatorname{var}_\rr(w_i) + \operatorname{var}_\rr(\varepsilon^*)$ (Lemma 3.1 of \cite{huang}). Specifically, after controlling for $V$, if the residual imbalance in $U$ is relatively small, it suggests that omitting $U$ in the estimation of the weights should result in $w_i$ that is similar to $w_i^*$, leading to $R^2_\varepsilon$ values close to 0. For instance, when $U$ is highly correlated with the observed effect modifiers, the estimated $w_i$ are expected to deviate minimally from the true weights $w_i^*$, resulting in a small $R_\varepsilon^2$. Conversely, if there is substantial residual imbalance in $U$, the weights $w_i$ will likely differ significantly from $w_i^*$, yielding large $R^2_\varepsilon$ values. In the extreme case where the weights $w_i$ are a constant, it implies no heterogeneity is explained by the observed weights, and $R^2_\varepsilon$ attains its maximum value of 1.

The $\rho_{\varepsilon, \tau}$ term quantifies the extent to which the unobserved $U$ explains treatment effect heterogeneity. As a correlation term, it naturally ranges from $-1$ to $1$, although a tighter bound can be derived using the recursive formula of partial correlation (See Section \ref{sec: corrbound} and Appendix B.2 of \cite{huang}). A large positive value of $\rho_{\varepsilon, \tau}$ suggests that individuals with larger individual treatment effects are overweighted, i.e., large $\tau_i$'s correspond to $w_i > w_i^*$ in the weighted estimator. Conversely, a large negative $\rho_{\varepsilon, \tau}$ indicates that these samples are underweighted. If $\rho_{\varepsilon, \tau}$ is close to zero, it implies that the residual imbalance in $U$ does not correlate with treatment effect heterogeneity, indicating $U$ is not an effect modifier and omitting $U$ is unlikely to introduce bias.

The $\sigma^2_\tau$ term represents the variance of the individual treatment effect and serves as a direct measure of treatment effect heterogeneity, which is inestimable as the potential outcomes cannot be observed simultaneously. When the amount of heterogeneity is substantial, generalization becomes more challenging, and even minor imbalances in $U$ can lead to significant bias. Conversely, when there is no treatment effect heterogeneity, adjustment for the distributions of effect modifiers is unnecessary, and the effect estimate from the collection of studies can be directly applied to the target population.

\cite{huang} further introduced several sensitivity tools to help interpret the sensitivity parameters. For example, the graphic contour plot visually represents the combinations \((R^2_\varepsilon, \rho_{\varepsilon, \tau})\) that would nullify the generalized effect estimates, providing an intuitive way for accessing robustness. The numerical robustness value \(RV_q\) quantifies the strength of an unobserved effect modifier $U$ needed to induce a bias equal to \(100 \times q \%\) of the generalized effect estimate, enabling researchers to assess the plausibility of such a modifier. The benchmarking tool minimum relative effect modifying strength (MREMS, previously referred to as minimum relative confounding strength (MRCS) in \cite{huang})) 
further allows researchers to calibrate their understanding of plausible parameter values by benchmarking the strength of the unobserved $U$ relative to observed modifiers. Since the proposed multi-study sensitivity framework adapts the same sensitivity parameters as in the single-trial case, these summary measures directly extend to the multi-study scenarios; for details see Section 4.1 of \cite{huang}.

\subsection{Parameter Estimation in the Multi-study Sensitivity Analysis}

Although the multi-study sensitivity framework inherits the same bias decomposition, sensitivity parameters, and analytic tools as in the single-trial analysis, application of those tools is more involved in this setting. Specifically, the estimation of the sensitivity parameter bounds and MREMS are more intricate because of the confounding structures and different treatment ratios across the studies. As a result, the naive estimators derived from the balanced single-trial scenario proposed in \cite{huang} are no longer applicable. In this section, we propose several estimators for sensitivity parameter bounds and formal benchmarking values that universally apply to any single- or multi-study generalization scenario.

\subsubsection{Estimating the upper bound of $\sigma_\tau^2$}
\label{sec: upperbound}

Since only one of the potential outcomes is observed for each individual, the sensitivity parameter $\sigma_\tau^2$ cannot be estimated. \cite{huang} suggests use of a sharp upper bound of $\sigma_\tau^2$, i.e., the theoretically maximally attainable $\sigma_\tau^2$ value, as a conservative estimate of the sensitivity parameter. We, instead, sharply bound $\sigma_\tau^2$ using the variances of the potential outcomes in the collection of studies by applying the Cauchy–Bunyakovsky–Schwarz inequality, such that
\begin{equation}
\label{varupper}
    \sigma_\tau^2 \leq \operatorname{var}_\rr(Y_i^1) + \operatorname{var}_\rr(Y_i^0) + 2 \sqrt{\operatorname{var}_\rr(Y_i^1) \operatorname{var}_\rr(Y_i^0) } \;,
\end{equation}
where equality is attained when the potential outcomes $Y_i^1, Y_i^0$ are perfectly uncorrelated, i.e., $\operatorname{cor}_\rr(Y_i^1, Y_i^0) = -1$.

It remains to estimate the potential outcome variances $\operatorname{var}_\rr(Y_i^a)$. In the single-trial scenario, these variances can be unbiasedly estimated as the sample variances of outcomes in the treatment and control groups, because of randomization. In the multi-trial scenario, the sample variance can be estimated via the pooled variance formula using the sample variance in each study
\[\widehat{\operatorname{var}}_\rr\left(Y^a\right) = \frac{\sum_{i = 1}^m (n_i - 1)\widehat{\operatorname{var}}_{S= i}\left(Y^a\right)}{\sum_{i = 1}^m (n_i - 1)} \;,\]
where $\widehat{\operatorname{var}}_{S= i}\left(Y^a\right)$ for $i = 1, \cdots, m$ are the sample variances of the potential outcomes estimated from each trial's treatment or control group. However, when generalizing from one or multiple observational studies, we need to adjust for confounding before estimating the variance. We propose a Hájek type inverse probability weighting (IPW) variance estimator, of the form
\begin{equation}
\label{hajekvar}
\begin{aligned}\widehat{\operatorname{var}}_{\mathcal{R}}^{\operatorname{IPW}}\left(Y^a\right) & = \widehat{\nu}_a - \widehat{\mu}^2_a\\
    & = \frac{\sum_{i \in \rr} I(A_i = a) Y_i^2 /P_{\mathcal{R}}\left(A_i=a \mid X_i, S_i\right)}{\sum_{i \in \rr} I(A_i = a) /P_{\mathcal{R}}\left(A_i=a \mid X_i, S_i\right)}\\
    & \quad \quad \quad \quad \quad \quad - \left[\frac{\sum_{i \in \rr} I(A_i = a) Y_i /P_{\mathcal{R}}\left(A_i=a \mid X_i, S_i\right)}{\sum_{i \in \rr} I(A_i = a) /P_{\mathcal{R}}\left(A_i=a \mid X_i, S_i\right)}\right]^2 \\
    & = \frac{\sum_\rr I(A_i = a) \left(Y_i - \widehat{\mu}_a\right)^2 / P_\rr(A_i = a \mid X_i, S_i)}{\sum_\rr I(A_i = a) / P_\rr(A_i = a \mid X_i, S_i)} \;.
\end{aligned}
\end{equation}
We provide the derivation of the estimator above in Appendix \hyperref[app:var]{C}. The estimator in equation (\ref{hajekvar}) can be viewed as the combination of two Hájek type estimators, namely, the difference between the estimator of the potential outcome second moment $\widehat{\nu}_a$, and the square of the estimator of the potential outcome mean $\widehat{\mu}^2$. We consider the Hájek type estimator rather than the commonly used Horvitz-Thompson type estimator for the following reasons. First, as noted by \cite{hajek}, when the study sample sizes are small, and/or when outcomes (or second moments of the outcomes) are weakly or negatively correlated with the propensity scores $P_\rr(A_i = a \mid X_i, S_i)$, the Hájek type estimator is typically more efficient compared to the Horvitz–Thompson estimator. More importantly, the proposed variance estimator always produces a positive estimate, unlike the Horvitz–Thompson approach, which can yield negative variance estimates under small sample sizes, especially when the estimated second moments are smaller than the square of potential outcome means due to large estimation variances.

Finally, although the sharp bound in equation (\ref{varupper}) can directly serve as a conservative estimate of $\sigma_\tau^2$, it may be further refined in practice. For details on how additional information or assumptions can reduce the variance upper bound, see Appendix A.4 of \cite{huang}. For general situations, we suggest using the sum of the potential outcome variances as a conservative estimate of $\sigma_\tau^2$, such that
\begin{equation}
    \label{eq: estsigma}
    \widehat{\sigma}_{\tau, \max}^2 \approx \widehat{\operatorname{var}}_{\mathcal{R}}\left(Y^1\right) + \widehat{\operatorname{var}}_{\mathcal{R}}\left(Y^0\right) \;.
\end{equation}
This approximation provides a conservative upper bound for common usage, as it would only be exceeded in the unlikely event that the potential outcomes are negatively correlated.

\subsubsection{Estimating the bounds for $\rho_{\varepsilon, \tau}$}
\label{sec: corrbound}
By following an argument identical to that in \cite{huang}, we can further bound $\rho_{\varepsilon, \tau}$ as
\begin{equation*}
    -\sqrt{1-\operatorname{cor}_{\mathcal{R}}^2\left(w_i, \tau_i\right)}\ \leq \rho_{\varepsilon, \tau} \leq \sqrt{1-\operatorname{cor}_{\mathcal{R}}^2\left(w_i, \tau_i\right)} \;.
\end{equation*}
In the single-trial scenario, \cite{huang} suggests that $\operatorname{cor}_{\mathcal{S}}\left(w_i, \tau_i\right)$ can be estimated as 
\begin{equation}
\label{eq:estcornaive}
    \widehat{\operatorname{cor}}_{\mathcal{R}}\left(w_i, \tau_i\right)=\frac{\widehat{\operatorname{cov}}_\mathcal{R}\left(w_i, Y_i^1\right)-\widehat{\operatorname{cov}}_\mathcal{R}\left(w_i, Y_i^0\right)}{\sqrt{\widehat{\sigma}_{\tau, \max}^2 \cdot \widehat{\operatorname{var}}_{\mathcal{R}}\left(w_i\right)}} \;,
\end{equation}
where $\widehat{\sigma}_{\tau, \max}^2$ is the conservative estimate from the previous subsection. Furthermore, due to randomization, $\widehat{\operatorname{cov}}_\mathcal{R}\left(w_i, Y_i^a\right)$ can be directly estimated using the observed generalization weights $w_i$ and outcomes in the treatment or control groups.
However, in the multi-study scenario, due to different treatment ratios across studies and possible confounding structures, the estimator in equation (\ref{eq:estcornaive}) no longer applies. To mitigate this, we propose a universal correlation estimator that applies to both single- and multi-study settings. Note that by Theorem \ref{thm2}, we can rewrite 
\begin{equation*}
\label{eq: cordecomp}
\begin{aligned}
    \operatorname{cov}_\mathcal{R}\left(w_i, \tau_i\right) = \ee_\rr(w_i \tau_i) - \ee_\rr(w_i )\ee_\rr( \tau_i) = \ee\left(\widehat{\tau}_W\right) - \ee_\rr( \tau_i) \;.
\end{aligned}
\end{equation*}
Thus, $\operatorname{cov}_\mathcal{R}\left(w_i, \tau_i\right)$ can be unbiasedly estimated as
\begin{equation*}
    \label{eq: estcor}
    \widehat{\operatorname{cov}}_{\mathcal{R}}\left(w_i, \tau_i\right)=\widehat{\tau}_W - \left(\widehat{\mu}_1 - \widehat{\mu}_0\right) \;,
\end{equation*}
where $\widehat{\mu}_1, \widehat{\mu}_0$ are the IPW estimators for potential outcome means. Finally, the estimated bounds of $\rho_{\varepsilon, \tau}$ are written as
\begin{equation}
    \label{eq:corbound}
    -\sqrt{1-\frac{\widehat{\operatorname{cov}}^2_{\mathcal{R}}\left(w_i, \tau_i\right)}{\widehat{\sigma}_{\tau, \max}^2 \cdot \widehat{\operatorname{var}}_{\mathcal{R}}\left(w_i\right)}}\ \leq \rho_{\varepsilon, \tau} \leq \sqrt{1-\frac{\widehat{\operatorname{cov}}^2_{\mathcal{R}}\left(w_i, \tau_i\right)}{\widehat{\sigma}_{\tau, \max}^2 \cdot \widehat{\operatorname{var}}_{\mathcal{R}}\left(w_i\right)}} \;.
\end{equation}

\subsubsection{Estimating Minimum Relative Effect Modifying Strength}
\label{sec: bench}
The MREMS helps researchers determine how strong an unobserved effect modifier must be, compared to a known modifier, to reduce the generalized effect estimate to zero. To estimate the MREMS for the unobserved effect modifier, according to Theorem 4.1 of \cite{huang}, we first estimate
\begin{equation}
    R_\varepsilon^{2-(j)} \equiv \operatorname{var}_\rr\left(\varepsilon_i^{-(j)}\right) /\operatorname{var}_\rr(w_i) \quad \text{and} \quad \rho_{\varepsilon, \tau}^{-(j)} \equiv \operatorname{cor}_\rr\left(\varepsilon_i^{-(j)}, \tau_i\right) \;,
\end{equation}
where $\varepsilon_i^{-(j)} \equiv w_i^{-(j)}  - w_i$ is the linear difference between the observed generalization weight $w_i$ and the generalization weight estimated with the $j$-th effect modifier omitted in the estimation, of the form
\[w_i^{-(j)} \equiv \frac{P(R_i = 1)}{P(R_i = 0)} \frac{P\left(R_i=0 \mid V_i^{-(j)}\right)}{P\left(R_i=1 \mid V_i^{-(j)}\right)} \;,\]
where $V_i^{-(j)}$ denotes all but the $j$-th observed effect modifiers for each $j = 1, \cdots, p$. 

For estimation, $R_\varepsilon^{2-(j)}$ can be directly estimated using the generalization weights in both single- and multi-study settings. When generalizing from a single trial, $\rho_{\varepsilon, \tau}^{-(j)}$ can be similarly estimated as in \eqref{eq:estcornaive} using outcomes from the treatment and control groups. In multi-study scenarios, however, estimating $\rho_{\varepsilon, \tau}^{-(j)}$ faces challenges similar to those in estimating the bounds of $\rho_{\varepsilon, \tau}$, which we claim can be addressed using analogous methods. To proceed, consider the generalizability estimator admitting $w_i^{-(j)}$
\begin{equation}
    \label{benchmark}
    \widehat{\tau}_{W}^{-(j)} \equiv\frac{1}{N_1}\sum_{i \in \mathcal{R}} w_i^{-(j)} \lambda_i \gamma_i A_i Y_i - \frac{1}{N_0}\sum_{i \in \mathcal{R}} w_i^{-(j)} \lambda_i \gamma_i (1 - A_i) Y_i \;,
    \end{equation}
where $\lambda_i,$ and $\gamma_i$ are as defined in (\ref{weight}). By an argument analogous to the proof of Theorem \ref{thm2}, the expectation of $\widehat{\tau}_{W}^{-(j)}$ is
 \[\ee\left(\widehat{\tau}_{W}^{-(j)}\right) = \ee_\rr\left(w_i^{-(j)} \tau_i\right) \;,\]
therefore we have
\begin{align*}
    \operatorname{cov}_\mathcal{R}\left(w_i^{-(j)}, \tau_i\right) = \ee_\rr\left(w_i^{-(j)} \tau_i\right) - \ee_\rr\left(w_i^{-(j)}\right)\ee_\rr( \tau_i) = \ee\left(\widehat{\tau}_{W}^{-(j)}\right) - \ee_\rr( \tau_i) \;.
\end{align*}
Thus, similar to Section \ref{sec: corrbound}, we can estimate $\rho_{\varepsilon, \tau}^{-(j)}$ as 
\begin{equation}
\label{eq: benchmarkest}
    \widehat{\rho}_{\varepsilon, \tau}^{-(j)} = \frac{\widehat{\operatorname{cov}}_\mathcal{R}\left(w_i^{-(j)}, \tau_i\right) -\widehat{\operatorname{cov}}_\mathcal{R}\left(w_i, \tau_i\right)}{\sqrt{\widehat{\sigma}_{\tau, \max}^2 \cdot \widehat{\operatorname{var}}_{\mathcal{R}}\left(\varepsilon_i^{-(j)}\right)}} = \frac{\widehat{\tau}_{W}^{-(j)} - \widehat{\tau}_{W}}{\sqrt{\widehat{\sigma}_{\tau, \max}^2 \cdot \widehat{\operatorname{var}}_{\mathcal{R}}\left(\varepsilon_i^{-(j)}\right)}} \;,
\end{equation}
which in fact applies to both single- and multi-study settings. Lastly, the MREMS for an unobserved effect modifier which has equivalent effect modifying strength to $V^{(j)}$ can be estimated as 
\begin{equation}
    \label{MREMS}
    \operatorname{MREMS}\left(X^{(j)}\right) = \frac{\widehat{\tau}_W}{ \widehat{\rho}_{\varepsilon, \tau}^{-(j)} \sqrt{\widehat{\sigma}_\tau^2 \cdot \widehat{R}_\varepsilon^{2-(j)} / \left(1 + \widehat{R}_\varepsilon^{2-(j)}\right)}  } \;.
\end{equation}
Furthermore, \cite{huang} further proposed MREMS$_\alpha$ (referred to in Huang et al. as MRCS$_\alpha$) to help researchers determine how strong an unobserved effect modifier must be, compared to a known modifier, to alter the statistical significance of the generalized effect estimate, which further incorporates the uncertainty of the generalized effect estimate in the sensitivity analysis.  MREMS$_\alpha$ can be similarly estimated as
\begin{equation}
    \label{MREMSalpha}
\operatorname{MREMS}_\alpha\left(X^{(j)}\right) = \frac{\mathrm{Minimal\;Bias\;Threshold}}{ \widehat{\rho}_{\varepsilon, \tau}^{-(j)} \sqrt{\widehat{\sigma}_\tau^2 \cdot \widehat{R}_\varepsilon^{2-(j)} / \left(1 + \widehat{R}_\varepsilon^{2-(j)}\right)}  } \;.
\end{equation}
For details on uncertainty quantification via bootstrap, and estimation of the Minimal Bias Threshold, refer to Appendix \ref{app: uncertainty}.

\subsection{Summary of the Sensitivity Framework}
\label{subsec: summary}
We now give a summary of the sensitivity framework that universally applies to any single- or multi-study generalization scenario.
\begin{itemize}
    \item[Step 1.] Estimate an upper bound of $\sigma^2_\tau$ as in Section \ref{sec: upperbound}.
    \item[Step 2.] Vary $\rho_{\varepsilon, \tau}$ by the bounds estimated as in Section \ref{sec: corrbound}.
    \item[Step 3.] Vary $R_{\varepsilon}^2$ across the range of $[0,1)$.
    \item[Step 4.] Compute adjusted PATE estimates with $\left(R_{\varepsilon}^2, \rho_{\varepsilon, \tau}, \sigma^2_\tau\right)$, and bootstrap to evaluate uncertainty following Appendix \ref{app: uncertainty}.
    \item[Step 5.] Generate the graphical contour plots, compute the numerical robustness value $RV_q$ as 
    \[
RV_q 
= \frac{1}{2}\Bigl(\sqrt{a_q^2 + 4 a_q} - a_q\Bigr),
\quad
\text{where}\quad
a_q = \frac{q^2 \cdot \hat\tau_W^2}{\sigma_{\tau,\max}^2 \cdot \mathrm{var}_R(w_i)} \;,
\] and estimate
MREMS and MREMS$_\alpha$ using estimators in Section \ref{sec: bench} to aid in the interpretation of the sensitivity parameters. 
\end{itemize}

\section{Hypothesis Testing for Unobserved Effect Modification}
\label{s:hypo}
In practice, we may consider an informal hypothesis testing approach to detect the presence of unobserved effect modifiers across multiple studies. Suppose we generalize effect estimates separately from each individual study to the target population. Assuming the positivity assumptions are satisfied, if the observed effect modifiers sufficiently adjust for the covariate imbalances between the studies and the target population, we would expect these generalized estimates to align closely across studies. Conversely, if the observed effect modifiers account for only a small portion of the covariate imbalance, it may be an indication that the generalizability assumption (A4) is substantially violated and suggests further sensitivity analysis is required.

We emphasize that the hypothesis testing method proposed here is exploratory rather than a formally consistent statistical test. Specifically, even with large sample sizes, the Type II error rate may not approach zero. This limitation implies that while the test can assess whether generalized effect estimates from different studies are statistically equivalent, it cannot distinguish whether these estimates truly reflect the PATE in the target population, or if all estimates share a systematic bias of the same magnitude and direction. Thus, this informal testing approach serves primarily as a heuristic to explore the plausibility of generalizability assumptions, rather than providing definitive conclusions.

The first step in constructing the hypothesis testing framework is to generalize from each of the individual studies separately. However, in practice, it is often the case that only a fraction of the individual studies have full covariate overlap with the target and can be directly generalized to the target. We suppose that there are $k$ out of $m$ studies that can directly generalize to the target, which we index as $s_{(1)}, s_{(2)}, ..., s_{(k)}$.  Since the proposed estimator in Theorem \ref{thm1} is a generalized version of the single-trial generalizability estimator, it can be separately applied to generalize from each of the $k$ studies to the target population, yielding estimates that we denote as $\widehat{\tau}_{W,s_{(1)}}, \widehat{\tau}_{W,s_{(2)}}, \ldots, \widehat{\tau}_{W,s_{(k)}}$. We also estimate the variances of the generalizability estimators, either through bootstrap or a sandwich type variance estimator, which we denote as $\widehat{\sigma}_{s_{(1)}}^2, \widehat{\sigma}_{s_{(2)}}^2, \ldots, \widehat{\sigma}_{s_{(k)}}^2$. 

We consider a traditional hypothesis testing framework, with a null hypothesis stating that all generalized effect estimates are equivalent and an alternative hypothesis that at least one differs significantly from the others. Formally, we define
\[
H_0: \widehat{\tau}_{W,s_{(1)}} = \widehat{\tau}_{W,s_{(2)}} = \cdots = \widehat{\tau}_{W,s_{(k)}}\;, \quad 
H_a: \exists\; i, j \in \{s_{(1)}, \cdots, s_{(k)}\}, \quad \widehat{\tau}_{W,s_{(i)}} \neq \widehat{\tau}_{W,s_{j)}} \;.
\]
We employ a Multivariate Wald test \citep{wald} to evaluate \(H_0\). Let \(C\) be the
\((k - 1) \times k\) contrast matrix where \(C_{i, 1} = 1\) and \(C_{i, i + 1} = -1\) for each $i = 1, \cdots, (k-1)$, taking the form
\[
\renewcommand{\arraystretch}{0.8} 
C =
\begin{bmatrix}
1 & -1 & 0 & 0 & \cdots & 0 \\
1 & 0 & -1 & 0 & \cdots & 0 \\
1 & 0 & 0 & -1 & \cdots & 0 \\
\vdots & \vdots & \vdots & \vdots & \ddots & \vdots \\
1 & 0 & 0 & 0 & \cdots & -1
\end{bmatrix}_{(k-1) \times k}.
\]
Define the vector of generalized effect estimates as \(\widehat{\tau}_{W, S} \equiv (\widehat{\tau}_{W,s_{(1)}}, \widehat{\tau}_{W,s_{(2)}}, \cdots, \widehat{\tau}_{W,s_{(k)}})\), and the corresponding covariance matrix as \(\Sigma_S \equiv \operatorname{Diag}(\widehat{\sigma}_{s_{(1)}}^2, \widehat{\sigma}_{s_{(2)}}^2, \cdots, \widehat{\sigma}_{s_{(k)}}^2)\). The off-diagonal entries of $\Sigma_S$ are zero because, under the assumed Bernoulli sampling scheme, each study’s samples are collected independently, so that when the covariate distribution in the target population sample is fully observed and not estimated, the corresponding generalized effect estimates are also independent. Under the null hypothesis, the large sample test statistic of the Multivariate Wald test is given by
\begin{equation}
H_0: \quad \left(C\widehat{\tau}_{W, S}\right)^T\left[C \Sigma_S C^T\right]^{-1} \left(C \widehat{\tau}_{W, S}\right) \to \chi^2_{k-1} \quad \text{as } n \to \infty \;,
\end{equation}
such that asymptotically, the test statistics should follow a $\chi^2$ distribution with $(k-1)$ degrees of freedom. 

Convention is to reject the null hypothesis at a significance level of $\alpha = 0.05$. However, in practice, limited sample sizes in individual studies can lead to large variances for the generalized estimators, thus reducing the power of the Wald test. To mitigate this issue, we recommend increasing the significance level $\alpha$ from 0.05 to 0.1 or 0.15, depending on the study sample sizes. Although this adjustment may raise the Type I error rate and lead to a more conservative test, our primary goal is to assess the robustness of the generalized estimator, making a trade-off between Type I and Type II errors acceptable. In other words, we advise direct generalization only when there is no moderate evidence rejecting the null. Otherwise, if the test is rejected or yields relatively small p-values, we suggest conducting additional sensitivity analyses or reconsidering the eligibility for generalization.

\subsection{Simulation}
\label{sec: sim}
To evaluate the power of the aforementioned test, we conduct the following simulations, which build on the simulations in \cite{casmeta}. We consider generalizing from three randomized controlled trials $(S = 1, 2, 3)$ to a target population $(S = 0)$, where each of the trials has the same sample size as the target population sample, varying from $500, 1000, 2000$ (i.e., $n_1 = n_2 = n_3 = n_0 = 500, 1000, 2000$). For each participant in the trials and target population sample, we generate three effect modifiers $X^{(1)}, X^{(2)}, X^{(3)}$ from a mean zero multivariate normal distribution with all marginal variances equal to $1$ and all pairwise correlations equal to $0.5$. We consider a logistic-linear model for the probability of selection into the combination of trials, such that
\[
R_i \sim \operatorname{Bernoulli}(p_i), \quad \text{where } \text{logit}(p_i) = \beta^T X_i \;,
\]
with $X_i = \left(1, X_i^{(1)}, X_i^{(2)}, X_i^{(3)}\right)$, $\beta = (\beta_0, \ln(1.25), \ln(1.25), \ln(1.25))$, and $\beta_0$ is computationally optimized to result in (on average) the desired sample size for the collection of trials. We then allocated trial participants to one of the three randomized trials using a multinomial logistic model, such that 
\begin{align*}
S_i = & 1, 2, 3 \mid  (X_i, R_i = 1) \sim \operatorname{Multinomial}(\pi_1, \pi_2, \pi_3)\;, \quad \text{where} \\
\pi_1 &= P[S_i = 1 \mid X_i, R_i = 1] = \frac{ \exp\left(\xi^T X_i\right)}{1 + \exp\left(\xi^T X_i\right) + \exp\left(\zeta^T X_i\right)} \;, \\
\pi_2 &= P[S_i = 2 \mid X_i, R_i = 1] =  \frac{\exp\left(\zeta^T X_i\right)}{1 + \exp\left(\xi^T X_i\right) + \exp\left(\zeta^T X_i\right)}\;, \\
\pi_3 &= P[S_i = 3 \mid X_i, R_i = 1] = 1 - \pi_1 - \pi_2\;,
\end{align*}
with $\xi = (\xi_0, \ln(1.25), \ln(1.25), \ln(1.25))$, $\zeta = (\zeta_0, \ln(1.5), \ln(1.5), \ln(1.5))$, and $\xi_0, \zeta_0$ are calculated to result in the desired sample sizes for the trials. The above parameter choices ensure that on average the positivity assumptions (A2, A4) are satisfied, with the marginal standardization mean difference (SMD) for each covariate between each trial and the target population sample being less than 1. We further assume a balanced design for each trial, such that
\[A_i \mid S_i = 1, 2, 3 \sim \operatorname{Bernoulli}(0.5) \;.\] 
Lastly, we generate potential outcomes as 
\[Y^a_i = 5 + X^{(1)}_i + X^{(2)} + kX^{(3)}_i + a\left[-5 - 2X^{(1)_i} - 2X^{(2)}_i - 2kX^{(3)}_i\right] + \varepsilon_i, \quad \varepsilon \sim N(0, 1) \;.\]
where $k = 1, 1.5$ is the scaling factor that controls the strength of $X^{(3)}$, resulting in a moderate or strong effect modifier. For each combination of $(n, k)$, we generate $1,000$ sets of trial and target population samples. 

We assume that $X^{(3)}$ is omitted in the estimation of the generalization, combination, and de-confounding weights, and apply the proposed estimator to generalize from each trial to the target population. We use entropy balancing \citep{Hainmueller2012-eb} to estimate the generalization weight, use the sample means of $A_i$ to estimate the treatment ratio in each trial and in the collection of trials, and use logistic regression to estimate the propensity score in the de-confounding weight. To estimate the variance, we use $1,000$ stratified bootstrap draws of each trial's treatment and control groups. We then apply the Multivariate Wald test and record the proportion of simulations in which the tests successfully identify that there is violation of the generalizability assumption A5 due to unobserved effect modification. 

We display the simulation results under different sample sizes, unobserved effect modifier strengths, and significance levels in Table \ref{tab1}. Larger sample sizes and stronger unobserved effect modifiers both lead to more successful identification, as expected. With a trial sample size $n = 1000$, the proposed test correctly identifies strong unobserved effect modification in $88.4\%$ of the simulations. Moreover, increasing the hypothesis testing significance level $\alpha$ appears beneficial for detecting unobserved effect modifications, particularly under realistic sample sizes where the variances of generalized weighted estimates can be large due to slower convergence rates. With $n = 1000$, the tests conducted at higher $\alpha$ levels correctly identifies moderate effect modification in over $80\%$ of the simulations, a substantial improvement compared to the conventional $\alpha = 0.05$ at $65\%$.


\begin{table}[ht]
\centering
\ra{0.75}  
\begin{tabular}{@{}ccccccc@{}}
\toprule
& \multicolumn{2}{c}{$\boldsymbol{\alpha = 0.05}$} 
& \multicolumn{2}{c}{$\boldsymbol{\alpha = 0.1}$} 
& \multicolumn{2}{c}{$\boldsymbol{\alpha = 0.15}$}\\
\cmidrule(r){2-3} \cmidrule(r){4-5} \cmidrule(r){6-7}
& \normalsize{$k = 1$} & \normalsize{$k = 1.5$}  
& \normalsize{$k = 1$} & \normalsize{$k = 1.5$} 
& \normalsize{$k = 1$} & \normalsize{$k = 1.5$}\\
\midrule
$\boldsymbol{n = 500}$  & 0.378  & 0.593  & 0.512  & 0.720  & 0.585  & 0.780 \\
$\boldsymbol{n = 1000}$ & 0.650  & 0.884  & 0.757  & 0.941  & 0.809  & 0.961 \\
$\boldsymbol{n = 2000}$ & 0.929  & 0.997  & 0.958  & 0.999  & 0.976  & 1.000 \\
\bottomrule
\end{tabular}
\caption{\footnotesize{Proportion of simulations in which the multivariate Wald test
         successfully identifies existence of an unobserved effect modifier.}}
\label{tab1}
\end{table}

\section{Application of the Sensitivity Analysis Framework to ECHO Cohort Data}
\label{s:echo}

The Environmental Influences on Child Health Outcomes (ECHO) Cohort is a nationwide consortium of 69 cohort sites including over 60,000 children residing across 50 states, the District of Columbia, and Puerto Rico. Each cohort site contributes a broad spectrum of exposures collected in both the parent cohort study and prospectively collected using a standardized protocol in ECHO \citep{echo}. ECHO leverages a standardized core protocol that measures five key outcome areas: pre-, peri-, and postnatal outcomes; neurodevelopment; airways; obesity; and positive health. This design enables rigorous, longitudinal analysis of how early-life environmental factors, from the prenatal period through adolescence, shape child health and development. By integrating such comprehensive data across diverse cohort sites, ECHO facilitates robust, large-scale studies on critical pediatric health questions, with the ultimate aim of informing effective interventions and prevention strategies. 

In a previous study, \cite{echoshs} investigated how well the effect of prenatal secondhand smoke (SHS) exposure on newborn birth weight could be generalized across the five largest eligible cohort sites in the ECHO Consortium (i.e., sites with exposure and outcome data and no inclusion or exclusion criteria directly affecting the outcome) using a targeted maximum likelihood generalizability estimator \citep{karatransport}. They selected the Healthy Start Study with participants recruited in Denver, Colorado as the target population given its minimal information bias (i.e., lowest risk of exposure misclassification or unmeasured confounding) and maximal covariate overlap with the other four cohort sites analyzed. Each of the four cohort sites was then individually transported to the Healthy Start Study cohort. Their findings suggested that these generalized effect estimates did not align with the causal effect directly estimated from the Healthy Start Study cohort, and that only a small to moderate portion of the observed variability could be attributed to differences in distributions among the measured socio-demographic characteristics.

In this application, we examine the same exposure–outcome relationship (SHS on birth weight) investigated by \cite{echoshs}, using the ECHO data, as a proof of concept for our proposed sensitivity framework. We stress that this analysis is not a definitive assessment of the ECHO cohort sites’ generalizability for this research question, as information bias, unmeasured confounding, and other potential sources of bias may persist, which our current method does not simultaneously address. Nevertheless, we hope that these findings, together with those of \cite{echoshs}, will provide deeper insights into how reliably the effect of SHS on newborn birth weight within ECHO can be generalized.


Exposure was defined as self-reported SHS during pregnancy, collected through prenatal questionnaires administered according to each cohort’s protocol. The outcome was birth weight, measured in grams and obtained from multiple sources across cohorts, including medical records, direct measurements by individual cohort study personnel, and self-reports \citep{echoshs}. For our study population, we began with all 54 ECHO cohort sites that recorded the relevant SHS exposures and newborn birthweight outcomes. We then excluded 28 cohort sites whose inclusion or exclusion criteria could directly influence exposure (e.g., cohorts restricted to active smokers) or outcomes (e.g., cohorts restricted by extreme birth weight or gestational age). We also exclude any participants who reported active maternal smoking during pregnancy. Although we aimed to include as many cohort sites as possible, we required each to have at least 150 participants to mitigate extreme de-confounding weights, leaving 14 cohorts eligible for generalization to Cohort~1. For additional information regarding these cohorts, refer to Appendix \hyperref[ap: cohort]{E}. 

We considered the same socio-demographic covariates as \cite{echoshs}, including self-reported infant sex at birth (Male, Female), infant race (White, Black, Asian, Multiple Race, Other Race), infant ethnicity (Hispanic vs. non-Hispanic), birthing parent age at delivery (in years), maternal pre-pregnancy body mass index (BMI measured continuously in \(\text{kg/m}^2\)), maternal education level (higher than high school, high school/General Educational Diploma (GED), some college, college degree, graduate degree), marital status (married, widowed/divorced, single), and family income (\(\geq\$30{,}000\) vs.\ \(<\$30{,}000\)). To avoid extreme weights, we combined Hawaiian and American Indian into the ``Other Race” category, given the small number of participants in these groups within the cohorts analyzed. Similarly as \cite{echoshs}, all covariates are considered as potential effect modifiers, and all covariates except for infant sex are evaluated as potential confounders. The characteristics for each cohort are summarized in Table \ref{tab:baseline}.
 
\begin{table}[ht!]
\centering
\footnotesize
\setlength{\tabcolsep}{3.5pt}
\renewcommand{\arraystretch}{0.75}
\resizebox{\textwidth}{!}{
\begin{tabular}{@{}lccccc@{}}
\toprule
\textbf{Characteristic} &
\begin{tabular}[c]{@{}c@{}}\textbf{Cohort Site 1}\\ \textit{(n = 1,149)}\end{tabular} &
\begin{tabular}[c]{@{}c@{}}\textbf{Cohort Site 2}\\ \textit{(n = 1,583)}\end{tabular} &
\begin{tabular}[c]{@{}c@{}}\textbf{Cohort Site 3}\\ \textit{(n = 1,538)}\end{tabular} &
\begin{tabular}[c]{@{}c@{}}\textbf{Cohort Site 4}\\ \textit{(n = 849)}\end{tabular} &
\begin{tabular}[c]{@{}c@{}}\textbf{Cohort Sites 2--15} \textit{(n = 10,698)}\end{tabular}\\
\midrule
\textbf{Birth weight [g; mean (SD)]} & \meansd{3,228.32}{504.21}  & \meansd{3,330.19}{526.86} & \meansd{3,265.12}{511.14} & \meansd{3,393.24}{472.93} & \meansd{3,328.29}{533.19} \\
\textbf{Maternal age [y; mean (SD)]} & \meansd{28.57}{6.20}  & \meansd{31.03}{5.15} & \meansd{32.78}{5.34} & \meansd{30.38}{4.61} & \meansd{31.41}{5.26} \\
\quad Missing & $-$ & $-$ & $-$ & $<$5 & $<$5 \\
\textbf{Prepregnancy BMI [kg/m$^2$; mean (SD)]} & \meansd{25.57}{6.11} & \meansd{27.53}{6.63} & \meansd{25.59}{5.63} & \meansd{27.86}{7.96} & \meansd{26.74}{6.72} \\
\quad Missing & $<$5 & $-$ & 275 & 12  & 764  \\
\addlinespace[2pt]
\textbf{Race [n (\%)]} & & & & & \\
\quad White               & 825 (71.9)    & 569 (42.4)  & 761 (53.6)  & 731 (86.6)& 6,151 (61.9) \\
\quad Black               & 139 (12.1)      & 118 (8.8)   & 98 (6.9)   & 7 (0.8)  & 1,319 (13.3) \\
\quad Asian               & 29 (2.5)        & 269 (20.1)  & 135 (9.5)    & 7 (0.8)& 684 (6.9)\\
\quad Multirace           & 53 (4.6)        & 300 (22.4)  & 100 (7.0)  & 59 (7.0)& 997 (10.0)\\
\quad Other race          & 102 (8.9)       & 85 (6.3)    & 327 (23.0) & 40 (4.7) & 781 (7.9) \\
\quad Missing             & $<$5            & 242         & 117     & 5    & 766 \\
\addlinespace[2pt]
\textbf{Hispanic [n (\%)]} & & & & & \\
\quad Yes                 & 346 (30.1)    & 780 (52.0) & 685 (51.5) & 37 (4.4) & 2,740 (27.1) \\
\quad No                  & 802 (69.9)  & 720 (48.0) & 645 (48.5) & 811 (95.6) & 7,363 (72.9) \\
\quad Missing             & $<$5                & 83         & 208    & $<$5    & 595 \\
\addlinespace[2pt]
\textbf{Sex [n (\%)]} & & & & & \\
\quad Female              & 546 (48.1)  & 785 (49.6) & 748 (48.7) & 420 (49.5) & 5,304 (49.6) \\
\quad Male                & 589 (51.9)  & 798 (50.4) & 787 (51.3) & 429 (50.5) & 5,391 (50.4) \\
\quad Missing             & 14                 & $-$         & $<$5   & $-$      & $<$5 \\
\addlinespace[2pt]
\textbf{Maternal education [n (\%)]} & & & & & \\
\quad $<$High school      & 146 (12.7)   & 34 (2.1)  & 161 (10.6) & 14 (2.1)& 444 (4.6) \\
\quad High school/GED     & 197 (17.1)   & 130 (8.2) & 248 (16.3) & 54 (8.3)& 1,100 (11.4)\\
\quad Some college        & 256 (22.3)  & 585 (37.0) & 232 (15.3) & 175 (26.8)& 2,568 (26.6) \\
\quad College degree      & 273 (23.8)  & 487 (30.8) & 383 (25.2) & 263 (40.2)& 2,990 (30.9)\\
\quad Graduate degree     & 277 (24.1)  & 346 (21.9) & 495 (32.6) & 148 (22.6)& 2,561 (26.5) \\
\quad Missing             & $-$                 & $<$5       & 19     & 195    & 1035 \\
\addlinespace[2pt]
\textbf{Marital status [n (\%)]} & & & & & \\
\quad Married             & 936 (81.9)  & 1,361 (86.0) & 1,264 (82.2) & 627 (92.5) & 8,201 (84.0) \\
\quad Widowed/divorced    & 22 (1.9)     & 29 (1.8)   & 8 (0.5)  & 10 (1.5)   & 344 (3.5) \\
\quad Single              & 185 (16.2)    & 192 (12.1) & 265 (17.2)& 41 (6.0) & 1,219 (12.5) \\
\quad Missing             & 6                & $<$5       & $<$5   & 171       & 934 \\
\addlinespace[2pt]
\textbf{Income [n (\%)]} & & & & & \\
\quad $<$USD \$30{,}000   & 146 (18.8)  & 175 (12.0) & 197 (18.1)& 95 (12.5)  & 1,350 (18.0) \\
\quad $\ge$USD \$30{,}000 & 630 (81.2) & 1,280 (88.0) & 892 (81.9)& 664 (87.5) & 6,170 (82.0)\\
\quad Missing             & 373                & 128        & 449    & 90     & 3,178 \\
\addlinespace[2pt]
\textbf{SHS exposure [n (\%)]} & & & & & \\
\quad Yes                 & 277 (24.1)   & 203 (12.8) & 185 (12.0) & 88 (10.4)& 2,169 (20.3) \\
\quad No                  & 872 (75.9)  & 1,380 (87.2) & 1,353 (88.2)& 761 (89.6) & 8,529 (79.7) \\
\bottomrule
\end{tabular}
}
\caption{\scriptsize{Data limited to analytical samples with complete exposure and outcome data and offspring with non-smoking mothers. - no data; BMI, body mass index; Cohort Site 1 (target), Healthy Start Study; Cohort Site 2, Pregnancy, Environment and Lifestyle study (PETALS); Cohort Site 3, NYU Children’s Health and Environment Study (NYU CHES); Cohort Site 4, Safe Passage Study; Cohort Sites 5-15, Combined summary of all other cohorts (See Appendix \ref{ap: cohort} for details. ECHO, Environmental influences on Child Health Outcomes; GED, General Educational Diploma; SD, standard deviation.}}
\label{tab:baseline}
\end{table}


Although exposure and outcome data were fully observed for all cohorts, some of the covariates had missing values. For illustrative purposes, we conducted single imputation via chained equations \citep{mice} separately within each cohort. After imputing the missing data, we first estimate the causal effect directly from the Healthy Start Study data with an IPW estimator, which yields an estimated effect on newborn birth weight of \(-44.94\)\,g [95\% CI \((-125.87,\,-33.90)\)]. We then applied the multi-study weighted estimator to generalize the effect estimate from the 14 eligible cohorts to the Healthy Start Study sample, which yields an estimate of \(\widehat{\tau}_W = -454.84\)\,g [95\% CI \((-754.96,\,-202.52)\)]. If little or no unobserved effect modification were present, we would expect the causal effect directly estimated from the target population (the Healthy Start Study) to align with the generalized effect estimates. However, we observe a substantial discrepancy between these estimates, with no overlap of the confidence intervals. From a practical standpoint, it is also unlikely that maternal SHS exposure would reduce birth weight by over 450 grams on average. Taken together, these observations suggest that the generalized effect estimate from the 14 cohort sites may not be reliable, and sensitivity analyses can help assess and interpret the robustness of the generalized results.

\subsection{Hypothesis Testing}

To ensure adequate power for the hypothesis test for the existence of unmeasured moderators, we focused on cohorts whose covariate distributions substantially overlapped with that of the target Healthy Start Study (i.e., each had a maximum marginal standardized mean difference (SMD) below 1 for all observed covariates), had minimal missingness for key covariates, and had comparatively large sample sizes (greater than 500 participants). Specifically, we computed generalized effect estimates from the Pregnancy Environment and Lifestyle Study (denoted as Cohort 2), the NYU Children’s Health and Environment Study (denoted as Cohort 3), and the Northern Plains Safe Passage Study (denoted as Cohort 4), as summarized in Table \ref{tab: echohypo}. Although some covariate data were missing in the target cohort and estimated through single imputation, according to Table \ref{tab:baseline}, the proportion of missingness was small. Therefore, we consider the generalized effect estimates to be approximately independent. A multivariate Wald test on these generalized effect estimates, with bootstrapped variances, yielded a p-value of 0.109. Under a conventional threshold of $\alpha = 0.05$, the null hypothesis would not be rejected. However, given the relatively small sample sizes of these cohorts and the consequent reduction in power, as suggested by our simulation results in Section \ref{s:hypo}, we view this as evidence suggestive of a violation of the generalizability assumption, which suggests further sensitivity analysis. 


\begin{table*}[hb!]
\label{tab: echohypo}
\centering
\ra{0.75} 
\begin{tabular}{@{}lcccc@{}}   
\toprule
& \textbf{Cohort Site Size} & \textbf{Max SMD}
& \textbf{Generalized Est} & \textbf{Bootstrapped Sd} \\
\midrule
\textbf{Cohort Site 2} & $n = 1583$ & 0.84 & -121.76 & 368.50 \\
\textbf{Cohort Site 3} & $n = 1538$ & 0.72 & 57.31   & 309.83 \\
\textbf{Cohort Site 4} & $n = 849$  & 0.73 & -1218.72 & 528.77 \\
\bottomrule
\end{tabular}
\caption{\footnotesize{Generalized effect estimates from Cohort Site 2, 3, and 4 to the Healthy Start Study sample,
         with bootstrapped variance estimates, where the estimate directly from Healthy Start Study is  \(-44.94\) [95\% CI \((-125.87,\,-33.90)\)]. The multivariate Wald test
         yields a p-value of 0.109.}}
\end{table*}


\subsection{Sensitivity Analysis}

We next deployed the graphical, numerical, and formal benchmarking tools to assess the robustness of the generalized effect estimates. 

We present a bias contour plot illustrating how different combinations of 
\(\bigl(R^2_\varepsilon, \rho_{\varepsilon,\tau}\bigr)\) can drive the generalized estimate of secondhand smoke exposure on newborn birth weight to alter the sign, to move it to zero, or to render it statistically insignificant in Figure 1.  The region beneath the solid blue line marks the full range of these parameter values sufficient to alter the sign of the estimated PATE. The solid blue line represents the combinations of \(\bigl(R^2_\varepsilon, \rho_{\varepsilon,\tau}\bigr)\) that will reduce the estimated PATE to zero. For example, an unobserved modifier accounting for half of the variation in the correctly specified generalization weights \(\bigl(R^2_\varepsilon = 0.5\bigr)\) and a moderate portion of the individual-level treatment 
effect \(\bigl(\rho_{\varepsilon,\tau} = -0.639\bigr)\) would nullify the estimate; likewise, a modifier explaining a substantial fraction of the generalization weights' variation 
\(\bigl(R^2_\varepsilon = 0.867\bigr)\) but only a modest fraction of the individual-level treatment 
effect \(\bigl(\rho_{\varepsilon,\tau} = 0.25\bigr)\) would have the same outcome. The region beneath the red dashed line captures the set of \(\bigl(R^2_\varepsilon, \rho_{\varepsilon,\tau}\bigr)\) that would make the PATE 
no longer statistically significant. As an illustration, when 
\(\bigl(R^2_\varepsilon, \rho_{\varepsilon,\tau}\bigr) = (0.25, -0.5)\), the bias-corrected estimate becomes \(-336.2\)\,g, with a $95\%$ bootstrapped CI \((-548.7,\,4.67)\), which now contains zero and thus is no longer statistically significant. 

\begin{figure}
    \centering
    \includegraphics[width = 0.9\linewidth]{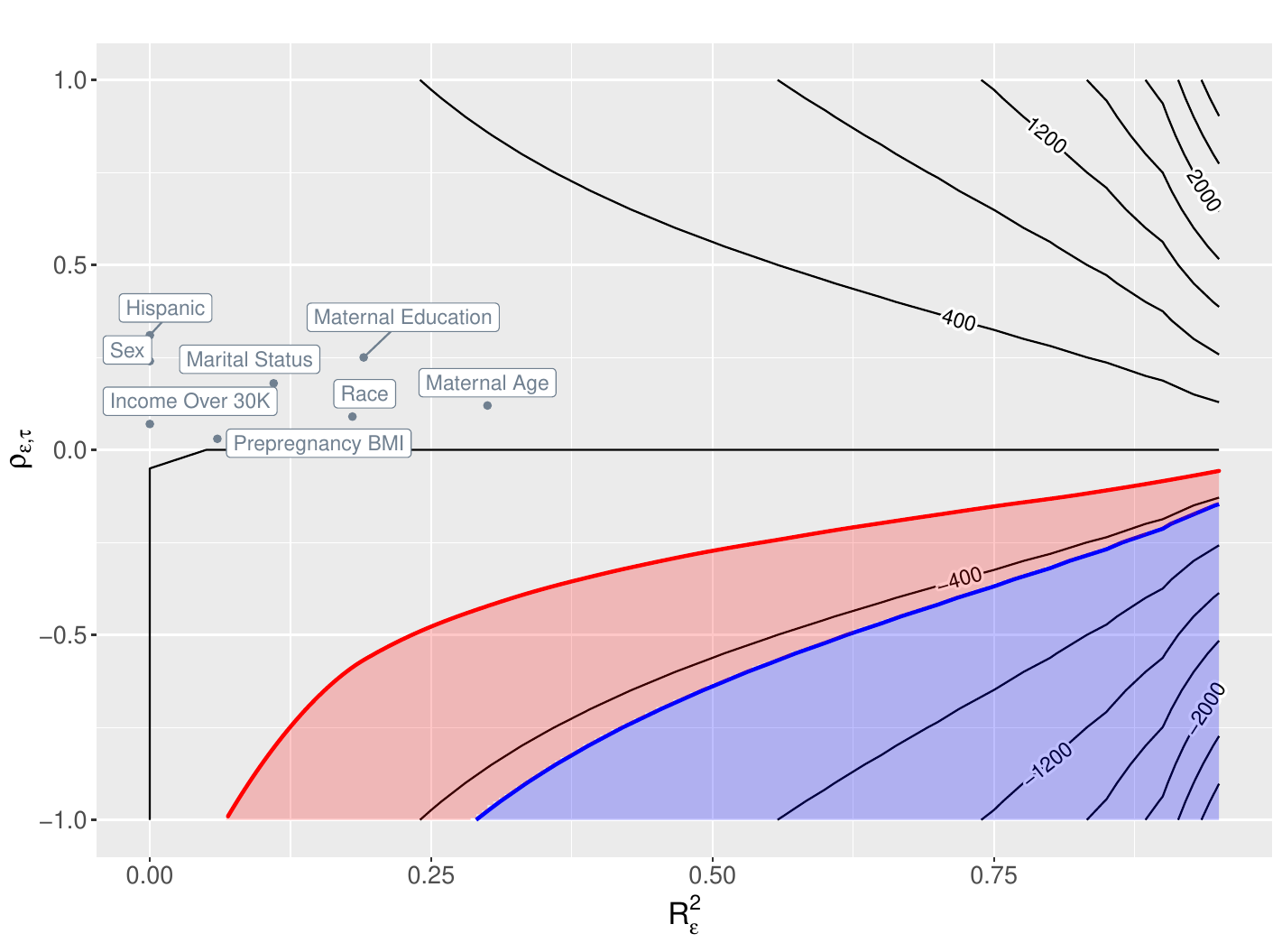}
    \caption{\footnotesize{Bias contour plot for generalizing to the Healthy Start Study sample. The inner region, bounded below the solid blue line, represents the combinations of $\bigl(R^2_\varepsilon, \rho_{\varepsilon,\tau}\bigr)$ produce enough bias to nullify or alter the sign of the generalized estimated effect. The outer region, bounded below the dashed red line,  represents those combinations that alter the statistical significance of the generalized effect estimate.}}
\end{figure}

For the numerical summaries, we first estimate the potential outcome variances using the H\'ajek type estimator in equation~\eqref{hajekvar}, then apply the conservative upper
bound in equation~\eqref{eq: estsigma}, obtaining an estimate of
\(\sigma_\tau = 767.1\). We then use the estimated \(\sigma_\tau^2\), generalized effect estimate $\widehat{\tau}_W$, and the estimated variance of the generalization weights $\widehat{\operatorname{var}}_\rr(w_i)$ to compute the numerical summaries for the sensitivity analysis. Specifically, following equation~(8) in
\cite{huang}, we obtain a robustness value of \(\operatorname{RV}_{q = 1} = 0.47\). This implies that omitting an effect modifier must account for 47\% of the variation in
both the individual-level treatment effect and 47\% of the variation in the correctly specified generalization weights, for the PATE to be reduced to zero. Whether 47\% is deemed large or small depends on whether researchers find it plausible that the omitted modifier could explain this proportion of variation in both the generalization weights and the treatment effect heterogeneity. Similarly, we estimate \(\operatorname{RV}_{\alpha = 0.05} = 0.24\), meaning that if omitting an effect modifier must account for 24\% of the variation in the generalization weights and the treatment effect heterogeneity, the estimated effect will no longer be
statistically significant.

Lastly, we perform formal benchmarking, and present the results in Table \ref{tab:MREMS}. For each observed effect modifier, the $\left(\widehat{R}^2_\varepsilon, \widehat{\rho}_{\varepsilon, \tau}\right)$ and the associated estimated bias represents the estimated sensitivity parameters and bias for if such modifier is omitted in the estimation of the weights. Furthermore, the MREMS used such estimated bias to benchmark how much relative modifying strength an omitted effect modifier $U$ must have compared to an observed effect modifier to reduced the estimated PATE to zero (MREMS) or no longer statistically significant (MREMS$_\alpha$). From the benchmarking results, we see that omitting an effect modifier with equivalent moderation strength to the covariates maternal education, maternal age, and marital status will result in the largest amount of bias. This is also reflected in the MREMS, which suggests that an unobserved effect modifier would have to be 5.33 -- 9.96 times stronger than any of the observed effect modifiers to reduce the estimated effect to zero. However, when accounting for uncertainty, we observe that an omitted moderator 2.35 times as strong as maternal education level would result in a statistically insignificant effect. As such, we conclude that while there is a large degree of robustness to an unobserved moderator reducing the point estimate to 0, there is some sensitivity to potential changes in the statistical significance of our estimated effect.


\begin{table}[ht]
\label{tab:MREMS}
\centering
\ra{0.75}  
\begin{tabular}{@{}lccccc@{}}
\toprule
\textbf{Variable}
 & $\boldsymbol{\widehat{R}^2_{\varepsilon}}$
 & $\boldsymbol{\widehat{\rho}_{\varepsilon,\tau}}$
 & $\boldsymbol{\widehat{\operatorname{Bias}}}$
 & \textbf{MREMS}
 & \textbf{MREMS$_\alpha$} \\
\midrule
\small\textbf{Sex}                & 0.00 & 0.24 & 6.42   & -70.83  & -31.26 \\
\small\textbf{Race}               & 0.18 & 0.09 & 31.65  & -14.37   & -6.34  \\
\small\textbf{Hispanic}           & 0.00 & 0.31 & 8.02   & -56.74  & -25.04 \\
\small\textbf{Maternal Age}       & 0.30 & 0.12 & 55.12  & -8.25   & -3.64  \\
\small\textbf{Income Over 30K}    & 0.00 & 0.07 & 0.95   & -476.77 & -210.43\\
\small\textbf{Maternal Education} & 0.19 & 0.25 & 85.42  & -5.33   & -2.35  \\
\small\textbf{Marital Status}     & 0.11 & 0.18 & 45.68  & -9.96   & -4.39  \\
\small\textbf{Prepregnancy BMI}   & 0.06 & 0.03 & 5.72   & -79.53  & -35.10 \\
\bottomrule
\end{tabular}
\caption{\footnotesize{Formal benchmarking results for generalizing from 14 cohort sites to the Healthy Start Study sample.}}
\end{table}


\section{Discussion}
Recent years have witnessed a growing number of collective research consortia, which pool independent collaborating studies to examine scientific questions. A fundamental methodological challenge in such designs is determining precisely to whom and how the pooled results generalize. Methodological advances suggest addressing this issue by adjusting the distribution of effect modifiers across the contributing studies and the target population. However, there is no guarantee that a complete set of effect modifiers is observed in the collection of studies and the target population, and the PATE estimate may be biased. 

In this work, we proposed a non-parametric sensitivity analysis framework for unobserved effect modifiers when generalizing from multiple studies to a target population. Building on \cite{huang}, our framework imposes no additional distributional or functional assumptions on the unobserved modifier, and inherent in the graphical, numerical, and formal benchmarking sensitivity analysis tools to aid better interpretation of the sensitivity parameters. We considered a causally interpretable, multi-study weighted generalizability estimator in \cite{casmeta}, modified to handle weaker overlap and generalizability assumptions while enhancing efficiency. We showed that the bias decomposition for the multi-study weighted estimator is identical to that of the single-trial scenario, thereby allowing direct application of the framework in \cite{huang}.  Moreover, we proposed universal estimation strategies for sensitivity parameter bounds and benchmarking parameters that applies to both single- and multiple- study generalization settings. We also suggested a hypothesis testing framework that uses the generalized estimates from each eligible study to detect unobserved effect modification. We showed with simulations that our proposed multivariate Wald test achieves good power when the bias in the studies varies in magnitude and directions, and we offered guidance on selecting appropriate significance levels under realistic sample sizes. 
We then applied the proposed sensitivity framework with hypothesis testing to the ECHO dataset to examine the generalizability of the effect of maternal prenatal secondhand smoke exposure on newborn birth weight.

There remain several important directions for future research. First, our analysis assumes that the unobserved effect modifier is not simultaneously a confounder. While this assumption may be justified in settings that combine randomized trials, it is less plausible in observational studies, where an effect modifier, by definition, is related to the outcome and may also act as a confounder. Extending sensitivity analyses to explicitly accommodate an unobserved covariate $U$ that serves both as an effect modifier and a confounder would therefore be an important advance.
Second, although the single-trial sensitivity framework in \cite{huang} extends to doubly robust augmented weighted estimators, no doubly robust estimator has yet been developed under assumptions A1–A5. Such an estimator could likely be derived by adapting the approach of Section 5 in \cite{casmeta}, and extending the multi-study sensitivity framework to incorporate doubly robust estimators would be of particular interest.
Third, the Multivariate Wald test has the limitation that failure to reject the null hypothesis does not establish equivalence of generalized effect estimates or exclude the possibility of unobserved effect modification; it merely reflects insufficient evidence to support the alternative. This limitation suggests that an equivalence testing framework may be more suitable. However, as detailed in Appendix \hyperref[ap: equiv]{D}, existing approaches do not provide a Wald-type multivariate equivalence test, highlighting the need for methodological development in this area.
Finally, in the present study, we used ECHO data primarily as a proof of concept rather than for substantive inference. Future work should pursue more comprehensive analyses, incorporating multiple imputation in conjunction with benchmarking values, addressing non-probability sampling in the target population \citep{Elliott2023-re}, and accounting for intercluster correlation in generalization.

\newpage
\section*{Acknowledges}
\textbf{ECHO Acknowledgments:} The authors wish to thank our ECHO Colleagues; the medical, nursing, and program staff; and the children and families participating in the ECHO cohort.\\~\\
\textbf{Funding Statement and Disclaimer:} The content is solely the responsibility of the authors and does not necessarily represent the official views of the National Institutes of Health.\\~\\
Research reported in this publication was supported by the Environmental influences on Child Health Outcomes (ECHO) Program, Office of the Director, National Institutes of Health, under Award Numbers U2COD023375 (Coordinating Center), U24OD023382 (Data Analysis Center), U24OD023319 with co-funding from the Office of Behavioral and Social Science Research (Measurement Core), U24OD035523 (Lab Core), ES0266542 (HHEAR), U24ES026539 (HHEAR Barbara O’Brien), U2CES026533 (HHEAR Lisa Peterson), U2CES0\allowbreak26542 (HHEAR Patrick Parsons, Kannan Kurunthacalam), U2CES030859 (HHEAR Manish Arora), U2CES030857 (HHEAR Timothy R. Fennell, Susan J. Sumner, Xiuxia Du), U2CES026555 (HHEAR Susan L. Teitelbaum), U2CES026561 (HHEAR Robert O. Wright), U2CES030851 (HHEAR Heather M. Stapleton, P. Lee Ferguson), UG3/UH3OD023251 (Akram Alshawabkeh), UH3OD023320 and UG3OD035546 (Judy Aschner), UH3OD023332 (Clancy Blair, Leonardo Trasande), UG3/UH3OD023253 (Carlos Camargo), UG3/UH3OD0\allowbreak23248 and UG3OD035526 (Dana Dabelea), UG3/UH3OD023313 (Daphne Koinis Mitchell), UH3OD023328 (Cristiane Duarte), UH3OD023318 (Anne Dunlop), UG3/UH3OD023279 (Amy Elliott), UG3/UH3OD023289 (Assiamira Ferrara), UG3/UH3OD023282 (James Gern), UH3OD023287 (Carrie Breton), UG3/UH3OD023365 (Irva Hertz-Picciotto), UG3/UH3OD0\allowbreak23244 (Alison Hipwell), UG3/UH3OD023275 (Margaret Karagas), UH3OD023271 and UG3O\allowbreak D035528 (Catherine Karr), UH3OD023347 (Barry Lester), UG3/UH3OD023389 (Leslie Leve), UG3/UH3OD023344 (Debra MacKenzie), UH3OD023268 (Scott Weiss), UG3/UH3OD\allowbreak023288 (Cynthia McEvoy), UG3/UH3OD023342 (Kristen Lyall), UG3/UH3OD023349 (Thomas O’Connor), UH3OD023286 and UG3OD035533 (Emily Oken), UG3/UH3OD023348 (Mike O’Shea), UG3/UH3OD023285 (Jean Kerver), UG3/UH3OD023290 (Julie Herbstman), UG3/ \allowbreak UH3OD023272 (Susan Schantz), UG3/UH3OD023249 (Joseph Stanford), UG3/UH3OD023305 (Leonardo Trasande), UG3/UH3OD023337 (Rosalind Wright), UG3OD035508 (Sheela Sathyanarayana), UG3OD035509 (Anne Marie Singh), UG3OD035513 and UG3OD035532 (Annemarie Stroustrup), UG3OD035516 and UG3OD035517 (Tina Hartert), UG3OD035518 (Jennifer Straughen), UG3OD035519 (Qi Zhao), UG3OD035521 (Katherine Rivera-Spoljaric), UG3OD035527 (Emily S Barrett), UG3OD035540 (Monique Marie Hedderson), UG3OD035543 (Kelly J Hunt), UG3OD035537 (Sunni L Mumford), UG3OD035529 (Hong-Ngoc Nguyen), UG3OD035542 (Hudson Santos), UG3OD035550 (Rebecca Schmidt), UG3OD035536 (Jonathan Slaughter), UG3OD035544 (Kristina Whitworth).\\~\\
\textbf{Role of Funder Statement:} The sponsor, NIH, participated in the overall design and implementation of the ECHO Program, which was funded as a cooperative agreement between NIH and grant awardees. The sponsor approved the Steering Committee-developed ECHO protocol and its amendments including COVID-19 measures. The sponsor had no access to the central database, which was housed at the ECHO Data Analysis Center. Data management and site monitoring were performed by the ECHO Data Analysis Center and Coordinating Center. All analyses for scientific publication were performed by the study statistician, independently of the sponsor. The lead author wrote all drafts of the manuscript and made revisions based on co-authors and the ECHO Publication Committee (a subcommittee of the ECHO Operations Committee) feedback without input from the sponsor. The study sponsor did not review or approve the manuscript for submission to the journal. \\~\\
\textbf{Data Availability Statement:} Select de-identified data from the ECHO Program are available through NICHD’s Data and Specimen Hub (DASH). Information on study data not available on DASH, such as some Indigenous datasets, can be found on the ECHO study DASH webpage.
We wish to thank our ECHO colleagues; the medical, nursing, and program staff; and the children and families participating in the ECHO cohorts.

\bibliographystyle{apalike}
\bibliography{reference}

\newpage
\appendix
\section{Proof of Theorem \ref{thm1}}
\label{app: thm1}
We first derive the identification formula in Theorem \ref{thm2}. By the law of iterated expectations, the PATE can be rewritten as 
\[\mathbb{E}(Y^1 - Y^0 \mid R = 0) = \mathbb{E}[\mathbb{E}(Y^1 - Y^0 \mid V, R = 0 ) \mid R = 0] \;.\]
By A5, for any $s^* \in S$, we have 
\begin{align*}
    \mathbb{E}(Y^1 - Y^0 \mid V, R = 0) &= \mathbb{E}(Y^1 - Y^0 \mid V, S = s^*) \\
    &= \mathbb{E}(Y^1 \mid V, S = s^*) - \mathbb{E}(Y^0 \mid V, S = s^*) \;.
\end{align*}
For each conditional expectation of the potential outcome mean, we apply again the law of iterated expectations,
\begin{align*}
    &\mathbb{E}(Y^a \mid V, S = s^*) = \ee[\ee (Y^a \mid X, S = s^*) \mid V, S = s^*] \\
    &\quad = \ee[\ee (Y^a \mid A = a, X, S = s^*) \mid V, S = s^*] \quad \tag*{(A2)}\\
    &\quad = \ee[\ee (Y \mid A = a, X, S = s^*) \mid V, S = s^*] \quad \tag*{(A1)} \\
    &\quad = \ee\left[\ee \left[\frac{I(A = a)Y}{P(A = a \mid X, S = s^*,R = 1)} \mid X, S = s^*\right] \mid V, S = s^*\right] \quad \tag*{(A4)}\\
    &\quad = \ee \left[\frac{I(A = a)Y}{P(A = a \mid X, S = s^*,R = 1)} \mid V, S = s^*\right]  \tag*{\text{(Since $V\subseteq X$)}}\;.
\end{align*}
Therefore,
\begin{align*}
    &\mathbb{E}(Y^1 - Y^0 \mid V, R = 0) \\
    &\quad  =\ee \left[\frac{I(A = 1)Y}{P(A = 1 \mid X, S = s^*,R = 1)} - \frac{I(A = 0)Y}{P(A = 0 \mid X, S = s^*,R = 1)} \mid V, S = s^*\right] \\
    &\quad  =\ee \left[\frac{I(A = 1)Y}{P(A = 1 \mid X, S,R = 1)} - \frac{I(A = 0)Y}{P(A = 0 \mid X, S ,R = 1)} \mid V, S = s^*\right] \;.
\end{align*}
Since this holds for every $s^* \in S$, we conclude that 
\begin{align*}
    &\mathbb{E}(Y^1 - Y^0 \mid V, R = 0) \\
    &\quad =\ee \left[\frac{I(A = 1)Y}{P(A = 1 \mid X, S,R = 1)} - \frac{I(A = 0)Y}{P(A = 0 \mid X, S ,R = 1)} \mid V, R = 1\right] \;.
\end{align*}
Therefore, the PATE can be identified as
{\small
\begin{align*}
    &\mathbb{E}(Y^1 - Y^0 \mid R = 0) \\
    =&\ee \left[\ee \left[\frac{I(A = 1)Y}{P(A = 1 \mid X, S,R = 1)} - \frac{I(A = 0)Y}{P(A = 0 \mid X, S ,R = 1)} \mid V, R = 1\right] \mid R = 0 \right] \\
    =&\frac{1}{P(R = 0)} \ee \left[  I(R = 0) \ee \left[\frac{I(A = 1)Y}{P(A = 1 \mid X, S,R = 1)} - \frac{I(A = 0)Y}{P(A = 0 \mid X, S ,R = 1)} \mid V, R = 1\right]\right] \\
    =& \frac{1}{P(R = 0)} \ee \left[  I(R = 0) \ee \left[\frac{1}{P(R = 1 \mid V)}\left[\frac{I(A = 1, R =1)Y}{P(A = 1 \mid X, S,R = 1)} - \frac{I(A = 0, R = 1)Y}{P(A = 0 \mid X, S,R = 1)}\right] \mid V\right]\right] \quad \tag*{(A3)}  \\
    =& \frac{1}{P(R = 0)} \ee \left[\ee \left[\frac{P(R = 0 \mid V)}{P(R = 1 \mid V)}\left(\frac{I(A = 1, R =1)Y}{P(A = 1 \mid X, S,R = 1)} - \frac{I(A = 0, R = 1)Y}{P(A = 0 \mid X, S,R = 1)}\right) \mid V\right]\right]   \\
    =& \frac{1}{P(R = 0)} \ee \left[\frac{P(R = 0 \mid V)}{P(R = 1 \mid V)}\left(\frac{I(A = 1, R =1)Y}{P(A = 1 \mid X, S,R = 1)} - \frac{I(A = 0, R = 1)Y}{P(A = 0 \mid X, S,R = 1)}\right) \right]  \;.
\end{align*}}
Now we derive the Horvitz-Thompson type weighted estimator using the identification formula. We can approximate the expectations in the identification formula with the sample means of study participants who respectively received the treatment and control.
Thus, the weighted estimator of the PATE is
\[
    \widehat{\tau}_W = \frac{1}{N_1}\sum_{i \in \mathcal{R}} w_i \lambda_i \gamma_i A_i Y_i - \frac{1}{N_0}\sum_{i \in \mathcal{R}} w_i \lambda_i \gamma_i (1 - A_i) Y_i \;,
\]
where $N_a$ denotes the total number of individuals receiving treatment $A = a$ in the collection of studies, the subscript $\mathcal{R}$ suggests the sum over all samples in the collection of studies,
\begin{equation*}
    \begin{aligned}
        w_i &\equiv \frac{P(R_i = 1)}{P(R_i = 0)} \frac{P(R_i=0 \mid V_i)}{P(R_i=1 \mid V_i)}\;, 
        \quad \lambda_i = \frac{P(A_i \mid R_i = 1)}{P(A_i \mid S_i, R_i = 1)}\;, \\
         &\quad \quad \quad \quad \quad \text{and} \quad \gamma_i \equiv \frac{P(A_i \mid S_i, R_i = 1)}{P(A_i \mid X_i, S_i, R_i=1)} \;.
    \end{aligned}
\end{equation*}

\section{Proof of Theorem \ref{thm2}}
\label{app: thm2}
    
Let the subscript $\mathcal{D}$ denote the sum is taken over all samples in the collection of studies and target population. We first write the expectation of the misspecified weighted estimator $\widehat{\tau}_{W}$ 
\begin{align*}
        \ee\left(\widehat{\tau}_{W}\right) & = \mathbb{E}\left(\frac{1}{N_1} \sum_{i \in \mathcal{R}} w_i \lambda_i \gamma_i A_i Y_i - \frac{1}{N_0} \sum_{i \in \mathcal{R}} w_i   \lambda_i \gamma_i (1 - A_i) Y_i \right) \\
        & = \mathbb{E}\left(\frac{1}{N_1} \sum_{i \in \mathcal{D}}w_i  \lambda_i \gamma_i A_i Y_i R_i- \frac{1}{N_0} \sum_{i \in \mathcal{D}} w_i   \lambda_i \gamma_i (1 - A_i) Y_i R_i \right) \\
        & = \mathbb{E}\left(\frac{1}{N_1} \sum_{i \in \mathcal{D}} w_i  \lambda_i \gamma_i A_i Y_i^1 R_i- \frac{1}{N_0} \sum_{i \in \mathcal{D}} w_i  \lambda_i \gamma_i (1 - A_i) Y_i^0 R_i \right) \tag*{(A1)}\\
        & = \frac{1}{N_1}\sum_{i \in \mathcal{D}}\mathbb{E}\left(  w_i \lambda_i \gamma_i A_i Y_i^1 R_i\right) - \frac{1}{N_0}\sum_{i \in \mathcal{D}} \mathbb{E} \left(w_i \lambda_i  \gamma_i (1 - A_i) Y_i^0 R_i \right) \tag*{(Linearity)}\\
        & = \frac{1}{N_1}\sum_{i \in \mathcal{D}} \mathbb{E}\left( w_i  \lambda_i \gamma_i A_i Y_i^1 R_i \mid A_i = 1, R_i = 1\right) P(A_i = 1, R_i = 1) \\
        & \quad \quad \quad - \frac{1}{N_0}\sum_{i \in \mathcal{D}}\mathbb{E}\left(  w_i  \lambda_i \gamma_i A_i Y_i^0 R_i \mid A_i = 0, R_i = 1\right) P(A_i = 0, R_i = 1) \\
        & = \frac{1}{n}\sum_{i \in \mathcal{D}} \mathbb{E}\left( w_i  \lambda_i \gamma_i A_i Y_i^1 R_i \mid A_i = 1, R_i = 1\right) \\
        & \quad \quad \quad \quad \quad  - \frac{1}{n}\sum_{i \in \mathcal{D}}\mathbb{E}\left(  w_i \lambda_i  \gamma_i A_i Y_i^0 R_i \mid A_i = 0, R_i = 1\right) \\
        & = \mathbb{E}\left( w_i \lambda_i  \gamma_i Y_i^1 R_i \mid A_i = 1, R_i = 1\right) - \mathbb{E}\left(  w_i \lambda_i  \gamma_i Y_i^0 R_i \mid A_i = 0, R_i = 1\right) \\
        & = \mathbb{E}_\mathcal{R}\left(w_i \lambda_i \gamma_i Y_i^1\mid A_i = 1\right) - \mathbb{E}_\mathcal{R}\left(  w_i \lambda_i \gamma_i Y_i^0 \mid A_i = 0\right) \;.
\end{align*}
For each treatment $A = a$, by the law of iterated expectation, we have
\begin{align*}
        &\mathbb{E}_\mathcal{R}\left(w_i \lambda_i \gamma_i Y_i^a\mid A_i = a\right) \\
        =& \sum_{s \in \mathcal{S}} \sum_x \mathbb{E}_\mathcal{R}\left(w_i \lambda_i \gamma_i Y_i^a\mid A_i = a, X_i = x, S_i = s\right) P_\mathcal{R}(X_i = x, S_i = s \mid A_i = a) \\
          \intertext{Since $\lambda_i$ and $\gamma_i$ are constants conditioning on $A_i = a, S_i = s, X_i = x$, we can move it outside of the expectation, }    
        =& \sum_{s \in \mathcal{S}} \sum_x \mathbb{E}_\mathcal{R}\left(w_i Y_i^a\mid A_i = a, X_i = x, S_i = s\right) \\
        & \quad \quad  \quad \quad \frac{P_\rr(A_i = a)}{P_\rr(A_i = a \mid S_i = s)} \frac{P_\rr(A_i = a \mid S_i = s)}{P_\rr(A_i = a \mid X_i = x, S_i = s)}  P_\mathcal{R}(X_i = x, S_i = s \mid A_i = a) \\
        =& \sum_{s \in \mathcal{S}} \sum_x \mathbb{E}_\mathcal{R}\left(w_i Y_i^a\mid A_i = a, X_i = x, S_i = s\right)P_\mathcal{R}(X_i = x, S_i = s) \\
         \intertext{Since $X$ are sufficient for adjusting the confounding within each study, and $w_i$ is a constant conditioning on $X_i = x$, by consistency we have}
        =& \sum_{s \in \mathcal{S}} \sum_x 
        \mathbb{E}_\mathcal{R}\left(w_i Y_i^a\mid X_i = x, S_i = s\right)P_\mathcal{R}(X_i = x, S_i = s) \tag*{(A2)} \\
        =& \mathbb{E}_\mathcal{R}\left(w_i Y_i^a\right) \;.
\end{align*}
Therefore,
\begin{equation*}
    \ee\left(\widehat{\tau}_{W}\right) = \mathbb{E}_\mathcal{R}\left(w_i Y_i^1\right) - \mathbb{E}_\mathcal{R}\left(w_i Y_i^0\right) = \mathbb{E}_\mathcal{R}\left(w_i \tau_i \right) \;.
\end{equation*}
We now write the expectation of the unbiased ideal weighted estimator $\widehat{\tau}_W^*$. From an identical argument, we have 
\[\ee\left(\widehat{\tau}_{W}^*\right) =  \mathbb{E}_\mathcal{R}\left(w_i^* \lambda_i^* \gamma_i^* Y_i^1\mid A_i = 1\right) - \mathbb{E}_\mathcal{R}\left(  w_i^* \lambda_i^* \gamma_i^* Y_i^0 \mid A_i = 0\right)\;.\]
Similarly, for each treatment $A = a$, by the law of iterated expectation, we have 
\begin{align*}
        &\mathbb{E}_\mathcal{R}\left(w_i^* \lambda_i^* \gamma_i^* Y_i^a\mid A_i = a\right) \\
        =& \sum_{s \in \mathcal{S}} \sum_{x, u} \mathbb{E}_\mathcal{R}\left(w_i^* \lambda_i^* \gamma_i^* Y_i^a\mid A_i = a, X_i = x, U_i = u, S_i = s\right) P_\mathcal{R}(X_i = x, U_i = u, S_i = s \mid A_i = a) \\
          \intertext{Since $\lambda_i^*$ and $\gamma_i^*$ are constants conditioning on $A_i = a, S_i = s, X_i = x, U_i = u$, we can move it outside of the expectation, }    
        =& \sum_{s \in \mathcal{S}} \sum_{x, u} \mathbb{E}_\mathcal{R}\left(w_i^* Y_i^a\mid A_i = a, X_i = x, U_i = u, S_i = s\right) \frac{P_\rr(A_i = a)}{P_\rr(A_i = a \mid S_i = s)} \\
        & \quad \quad  \quad \quad \quad \quad \frac{P_\rr(A_i = a \mid S_i = s)}{P_\rr(A_i = a \mid X_i = x, U_i = u, S_i = s)}  P_\mathcal{R}(X_i = x, U_i = u, S_i = s \mid A_i = a) \\
        =& \sum_{s \in \mathcal{S}} \sum_{x, u} \mathbb{E}_\mathcal{R}\left(w_i^* Y_i^a\mid A_i = a, X_i = x, U_i = u, S_i = s\right)P_\mathcal{R}(X_i = x, U_i = u, S_i = s) \\
         \intertext{Since $X$ are sufficient for adjusting the confounding within each study, and $w_i$ is a constant conditioning on $X_i = x, U_i = u$, by consistency we have}
        =& \sum_{s \in \mathcal{S}} \sum_{x, u} 
        \mathbb{E}_\mathcal{R}\left(w_i^* Y_i^a\mid X_i = x, U_i = u, S_i = s\right)P_\mathcal{R}(X_i = x, U_i = u, S_i = s) \tag*{(A2)} \\
        =& \mathbb{E}_\mathcal{R}\left(w_i^* Y_i^a\right) \;.
\end{align*}
Therefore, since the estimator using the ideal weights are unbiased, we have
\begin{equation*}
    \ee\left(\widehat{\tau}_{W}^*\right) = \mathbb{E}_\mathcal{R}\left(w_i^* Y_i^1\right) - \mathbb{E}_\mathcal{R}\left(w_i^* Y_i^0\right) = \mathbb{E}_\mathcal{R}\left(w_i^* \tau_i \right) = \tau \;.
\end{equation*}
Thus, the bias of the multi-study weighted estimator when omitting $U$ in the estimation of the weights is 
\begin{align*}
    \operatorname{Bias}(\widehat{\tau}_W)& = \ee \left(\widehat{\tau}_W\right) - \tau \\
    & = \ee \left(\widehat{\tau}_W\right) - \ee \left(\widehat{\tau}_W^*\right)  \\
    & =  \mathbb{E}_\mathcal{R}\left(w_i \tau_i \right) -  \mathbb{E}_\mathcal{R}\left(w_i^* \tau_i \right) \\
    & =  \mathbb{E}_\mathcal{R}\left(\varepsilon_i \tau_i \right) \;.
\end{align*}
Since $\ee(w_i) = \ee(w_i^*) = 1$, we have $\ee(\varepsilon_i) = 0$, thus
\[\operatorname{Bias}(\widehat{\tau}_W) =  \mathbb{E}_\mathcal{R}\left(\varepsilon_i \tau_i \right) =  \mathbb{E}_\mathcal{R}\left(\varepsilon_i \tau_i \right) - \ee_\rr(\varepsilon_i) \ee_\rr(\tau_i) = \operatorname{cov}_\rr(\varepsilon_i, \tau_i) \;.\qed\] 

\section{Estimation of Potential Outcome Variance}
\label{app:var}
In the derivation the Hajek type IPW variance estimator, we will use a different notation for the potential outcomes, $Y_i(a)$, for clarity. The potential outcome variance in the collection of studies can be written as 
\begin{equation*}
    \operatorname{var}_\rr \left[Y_i(a)\right]= \ee_\rr \left[Y_i(a)^2\right] -  \left[\ee_\rr \left[Y_i(a)\right]\right]^2 \;.
\end{equation*}
We now identify the potential outcome variance by parts. First, by law of iterated expectations, 
\begin{align*}
        \ee_\rr \left[Y_i(a)^2\right] &= \ee_\rr \left[\ee_\rr\left[Y_i(a)^2 \mid X_i, S_i\right]\right] \\
        & = \ee_\rr \left[\ee_\rr\left[Y_i(a)^2 \mid A_i = a, X_i, S_i\right]\right] \quad \tag*{(A2)}\\
        & = \ee_\rr \left[\ee_\rr\left[Y_i^2 \mid A_i = a, X_i, S_i\right]\right] \quad \tag*{(A1)}\\
        & = \ee_\rr\left[\ee_\rr\left(\frac{I(A_i = a)Y_i^2}{P_\rr(A_i = a \mid X_i, S_i)} \mid X_i, S_i\right)\right] \\
        & = \ee_\rr\left[\frac{A_iY_i^2}{P_\rr(A_i = a \mid X_i, S_i)}\right]    \;.
\end{align*}
Thus, the corresponding Hajek type IPW estimator is 
\begin{equation*}
    \widehat{\nu}_a \equiv \frac{\sum_\rr A_i Y_i^2 / P_\rr(A_i = a \mid X_i, S_i)}{\sum_\rr A_i / P_\rr(A_i = a \mid X_i, S_i)} \;.
\end{equation*}
Similarly, by law of iterated expectations,
\begin{align*}
    \ee_\rr\left[Y_i(a)\right] &= \ee_\rr[\ee_\rr(Y_i(a) \mid X_i, S_i)] \\
    & = \ee_\rr[\ee_\rr(Y_i(a) \mid A_i = a, X_i, S_i)] \tag*{(A2)}\\
    & = \ee_\rr[\ee_\rr(Y_i \mid A_i = a, X_i, S_i)] \tag*{(A1)}\\
    & = \ee_\rr\left[\ee_\rr\left(\frac{I(A_i = a)Y_i}{P_\rr(A_i = a \mid X_i, S_i)} \mid X_i, S_i\right)\right] \\
    & = \ee_\rr\left[\frac{A_iY_i}{P_\rr(A_i = a \mid X_i, S_i)}\right]    \;.
\end{align*}
Thus, the corresponding Hájek type IPW estimator is 
\begin{equation*}
    \widehat{\mu}_a \equiv \frac{\sum_\rr A_i Y_i / P_\rr(A_i = a \mid X_i, S_i)}{\sum_\rr A_i / P_\rr(A_i = a \mid X_i, S_i)} \;.
\end{equation*}
Finally, the Hájek type potential outcome variance estimator can be written as 
\[\widehat{\operatorname{var}}_\rr^{\operatorname{IPW}}(Y^a) \equiv \widehat{\nu}_a - \widehat{\mu}_a^2 \;. \qed\]

\section{Accounting for Uncertainty in the Sensitivity Tools}
\label{app: uncertainty}

Following Appendix A.5 of \cite{huang}, we can use the percentile bootstrap framework proposed in \cite{uncer1} and \cite{uncer2} to assess uncertainty in the contour plot and calculating MREMS$_\alpha$. 

\begin{itemize}
    \item[Step 1.] Estimate $\sigma^2_{\tau,\max}$ and fix some \(\bigl(R^2_{\varepsilon}, \rho_{\varepsilon,\tau}\bigr)\).
Generate $B$ bootstrap samples independently for each study, thereby creating a collection of $B$ bootstrap samples across all studies.
    \item[Step 2.] For each collection of bootstrap sample \(b = 1, \dots, B\):
    \begin{itemize}
  \item Estimate generalization weights \(\widehat{w}^{(b)}_i\) and the point estimate \(\widehat{\tau}^{(b)}_W\).

  \item Using the fixed \(\sigma^2_{\tau, \max}\) value, calculate
    \(
       \widehat{\operatorname{cor}}_{\rr, b}\bigl(\widehat{w}^{(b)}_i,\tau_i\bigr) \text{ and }
       \widehat{\operatorname{var}}_{\rr, b} \bigl(\widehat{w}^{(b)}_i\bigr),
    \)
    where the subscript \(b\) denotes the quantity calculated over the \(b\)-th bootstrap sample.

  \item Using the bootstrapped quantities, calculate the adjusted weighted estimator for the \(b\)-th bootstrap:
    \[
      \widehat{\tau}^{*(b)}_{W} \bigl(R^2_{\varepsilon},\rho_{\varepsilon,\tau},\sigma^2_{\tau, \max}\bigr) \equiv
      \widehat{\tau}^{(b)}_{W}
      - \operatorname{Bias} \Bigl(\widehat{\tau}^{(b)}_{W}\,\bigm|\,
        R^2_{\varepsilon}, \rho_{\varepsilon,\tau}, \sigma^2_{\tau, \max}\Bigr) \;,
    \]
    using the bias decomposition formula in equation \eqref{biasformula}.
\end{itemize}
\item[Step 3.] From the \(B\) bootstrapped optimal bounds, estimate the \(\alpha/2\) and
\(1-\alpha/2\) percentiles of the minima and maxima values, respectively, to obtain valid confidence intervals:
\[
  \mathrm{CI}\bigl(\alpha,R^2_{\varepsilon}, \rho_{\varepsilon,\tau}\bigr)
  =
  \Bigl[
    Q_{\alpha/2}\Bigl(\widehat{\tau}^{*(b)}_{W}\bigl(R^2_{\varepsilon}, \rho_{\varepsilon,\tau}, \sigma^2_{\tau, \max}\bigr)\Bigr),
    \;
    Q_{1-\alpha/2}\Bigl(\widehat{\tau}^{*(b)}_{W}\bigl(R^2_{\varepsilon}, \rho_{\varepsilon,\tau}, \sigma^2_{\tau, \max}\bigr)\Bigr)
  \Bigr].
\]
\item[Step 4.]  Conduct a grid search over all combinations of 
\(\bigl(R^2_{\varepsilon}, \rho_{\varepsilon,\tau}\bigr)\), with 
\(R^2_{\varepsilon} \in (0,1)\) and \(\rho_{\varepsilon,\tau} \in (-1,1)\). For each $R^2_{\varepsilon}$, record the pair of \(\bigl(R^2_{\varepsilon}, \rho_{\varepsilon,\tau}\bigr)\) that changes the significance of adjusted weighted estimator based on the valid confidence interval $\mathrm{CI}\bigl(\alpha,R^2_{\varepsilon}, \rho_{\varepsilon,\tau}\bigr)$. The set of recorded $\mathrm{CI}\bigl(\alpha,R^2_{\varepsilon}, \rho_{\varepsilon,\tau}\bigr)$ constitutes the border of the contour plot's outer significance region. 
\item[Step 5.] For the recorded set of \(\bigl(R^2_{\varepsilon}, \rho_{\varepsilon,\tau}\bigr)\), use the bias decomposition formula in \eqref{biasformula} to compute
\[\mathrm{Minimal\;Bias\;Threshold} \equiv \underset{\text{Recorded } (R^2_{\varepsilon}, \rho_{\varepsilon,\tau})}{\operatorname{argmin}} \operatorname{Bias}\bigl(R^2_{\varepsilon}, \rho_{\varepsilon,\tau}, \sigma^2_{\tau, \max}\bigr) \;.\]
The estimated Minimal Bias Threshold is then used to compute the MREMS$_\alpha$ according to equation \eqref{MREMSalpha}.
\end{itemize}
 
\section{ECHO Cohort Information}
\label{ap: cohort}
In our data analysis, we treat the Healthy Start Study \citep{Sauder2016-yb} as the target cohort study and generalize findings from 14 eligible ECHO cohort studies to this target. Details of the target cohort study are provided in Table \ref{tab:echo_cohorts}. For the hypothesis testing in Section~\ref{s:hypo}, we focus on four ECHO cohort studies whose covariate distributions substantially overlap with those of the Healthy Start Study, have minimal missingness in key covariates, and have comparatively large sample sizes.

\begin{table}[ht!]
\centering
\ra{1.0} 
\setlength{\tabcolsep}{6pt} 
\footnotesize 
\begin{tabularx}{\linewidth}{@{}Ycc@{}}
\toprule
\textbf{ECHO Cohort Sites} & \textbf{Hypothesis Testing} & \textbf{Generalization} \\
\midrule
Healthy Start Study \textbf{(Target Cohort)} \citep{Sauder2016-yb} & \checkmark & \checkmark \\
\addlinespace 
Pregnancy Environment and Lifestyle Study (PETALS) \citep{Zhu2017-gh} & \checkmark & \checkmark \\
\addlinespace
The NYU Children's Health and Environment Study (NYU CHES) \citep{Trasande2020-wh} & \checkmark & \checkmark \\
\addlinespace
Safe Passage Study \citep{Dukes2014-ci} &  \checkmark & \checkmark \\
\addlinespace
Wayne County Health Environment Allergy and Asthma (WHEALS) \citep{Aichbhaumik2008-os} & & \checkmark \\
\addlinespace
New Hampshire Birth Cohort Study (NHBCS) \citep{Gilbert-Diamond2011-ox, Madan2016-mz, Gilbert-Diamond2016-sj, Signes-Pastor2021-lb} & & \checkmark \\
\addlinespace
ECHO in Puerto Rico \citep{Manjourides2020-pp, Ferguson2019-ki} &  & \checkmark \\
\addlinespace
The Kaiser Permanente Northern California research program (KPRB) \citep{Hedderson2016-ni} &  & \checkmark \\
\addlinespace
The Global Alliance to Prevent Prematurity and Stillbirth (GAPPS) \citep{Paquette2018-zp} &  & \checkmark \\
\addlinespace
Project Viva \citep{Oken2015-hz} &  & \checkmark \\
\addlinespace
Michigan Archive for Research in Child Health (MARCH) \citep{Elliott2025-jo} &  & \checkmark \\
\addlinespace
Illinois Kids Development Study (IKIDS) \citep{Eick2021-ph} &  & \checkmark \\
\addlinespace
Chemicals in Our Bodies (CIOB) \citep{Morello-Frosch2016-au} &  & \checkmark \\
\addlinespace
PRogramming of Intergenerational Stress Mechanisms (PRISM) \citep{Brunst2014-og} &  & \checkmark \\
\addlinespace
First 1000 Days Study \citep{Hazrati2016-cz} &  & \checkmark \\
\addlinespace
\bottomrule
\end{tabularx}
\caption{\footnotesize ECHO cohort sites and their use in hypothesis testing and generalization.}
\label{tab:echo_cohorts}
\end{table}

\section{Equivalence Testing for Detecting Unobserved Effect Modification}
\label{ap: equiv}

A limitation of the Multivariate Wald test is that failing to reject the null hypothesis does not establish the equivalence of the generalized effect estimate or rule out unobserved effect modification. Rather, it merely indicates insufficient evidence to support the alternative hypothesis. This issue is especially concerning when generalizing from studies with limited sample sizes, where a non-rejected null does not guarantee that the generalized effect estimate is unbiased with respect to unobserved effect modification. 

An equivalence testing framework may be more suitable for our purpose, which serves as the conceptual inverse of the traditional hypothesis testing framework. In this approach, an equivalence interval $(-\phi, \phi)$ is first defined, analogous to the traditional testing threshold $\alpha$, representing the range of differences within which the quantities are considered equivalent. The equivalence null hypothesis assumes that the quantities under comparison are not equivalent, such that the difference between quantities falls outside of the interval. Conversely, the equivalence alternative hypothesis instead assumes that the quantities are equivalent, with the difference lying within the interval. Consequently, rejecting the null implies that the quantities under comparison are considered equivalent. This is particularly appealing to our purposes, as failing to reject the null signals insufficient evidence for equivalence, thereby recommending further sensitivity analysis.


Unfortunately, most existing equivalence tests are ANOVA type tests, which are only designed to assess the equivalence of means across $m$ groups of independent samples, and are incompatible with the estimators we used for the PATE under asymptotical variances. To our knowledge, there is no direct analogue of the Multivariate Wald test in the equivalence testing literature. Nevertheless, we note that $(\widehat{\tau}_{W,1}, \widehat{\tau}_{W,2}, \cdots, \widehat{\tau}_{W,m})$ can be rewritten as
\begin{equation*}
    \widehat{\tau}_{W,s} \equiv \frac{1}{n_s} \sum_{S_i = s}\left[\frac{w_i A_i Y_i}{P(A_i = 1 \mid X_i = x, S_i = s)} - \frac{w_i (1 - A_i) Y_i}{P(A_i = 0 \mid X_i = x, S_i = s)}\right] \;,
\end{equation*}
which, for the sake of analysis, can be approximated as sample means for equivalence testing, although we admit the approximated ``samples" 
\[\Delta_{i} \equiv \frac{w_i A_i Y_i}{P(A_i = 1 \mid X_i = x, S_i = s)} - \frac{w_i (1 - A_i) Y_i}{P(A_i = 0 \mid X_i = x, S_i = s)}\]
are not independent due to correlations in the generalization weights and propensity scores. We consider the Wellek-Welch test proposed in \cite{koh}, which allows for the equivalence testing of $m$ independent groups with different sample sizes and variances. This method considers a test statistics $\psi^2$, which follows a scaled non-central $F$ distribution adapted from the traditional Welch procedure \citep{Welch}. The pair of hypothesis are expressed as
\[\widetilde{H}_0: \psi^2 \geq \phi^2, \quad \text{and} \quad \widetilde{H}_a: \psi^2 < \phi^2 \;,\]
where the alternative hypothesis can be heuristically interpreted as the normalized difference between any pair of sample means lying within a $k$-dimensional ball of radius $\varepsilon$, therefore are considered as equivalent. Common choices for $\varepsilon$ are 0.25 or 0.5, depending on if the test is rather strict or comparatively weak \citep{wellekbook}. For the exact form of the test statistic and the corresponding degrees of freedom, we omit them here and refer readers to \cite{koh}.

We conduct the same simulation as described in Section \ref{sec: sim}, varying for different sample sizes, modifier strength, and sizes of equivalence interval. We display the results in Table We observed that when using conventional equivalence intervals ($\phi = 0.25 \text{ or }0.5$), the test had no power to detect unobserved effect modification, likely because our samples are not independent. We then considered narrower intervals ($\phi = 0.1 \text{ or }0.01$) to make the test more inclined to deem the samples not equivalent, thereby increasing its power for our purpose. Surprisingly, at $\phi = 0.1$, the test’s conclusions depended heavily on sample size: it was more likely to reject equivalence under smaller sample sizes yet frequently suggested equivalence with larger samples.

Further simulations with independent sample means reveals that such phenomenon typically arises when the true normalized difference between quantities under comparison lies near the boundary of the equivalence interval. Under these borderline conditions, although the test will asymptotically reject the null, it consistently yields false positives in smaller samples.
Simulations further suggests two ways to avoid such irregular behavior: either reduce $\phi$ or increase the sample sizes. In practice, however, researchers rarely have the option to enlarge their samples, leaving adjustments to $\phi$ the only choice. Yet identifying a single $\phi$ that is neither overly conservative nor prone to false positives proved infeasible, given our relatively small, correlated samples (which further diminish the effective sample size). In other words, for any given $\phi$, we are unable to distinguish if the rejection of equivalence is because of $(1)$ the equivalence interval is overly narrow, $(2)$ a false positive arising from insufficient sample sizes, or $(3)$ genuine unobserved effect modification. 
 
These findings indicate that the Wellek–Welch test may not be suitable for detecting unobserved effect modification when the assumption of independent samples does not hold. We therefore conclude that, in multi-study settings, one should rely on the Multivariate Wald test for detecting unobserved effect modification.


\begin{table}[ht]
\centering
\ra{0.75}  
\begin{tabular}{@{}ccccccccc@{}}
\toprule
& \multicolumn{2}{c}{$\boldsymbol{\varepsilon = 0.5}$}
& \multicolumn{2}{c}{$\boldsymbol{\varepsilon = 0.25}$}
& \multicolumn{2}{c}{$\boldsymbol{\varepsilon = 0.1}$}
& \multicolumn{2}{c}{$\boldsymbol{\varepsilon = 0.01}$}\\
\cmidrule(r){2-3}\cmidrule(r){4-5}\cmidrule(r){6-7}\cmidrule(r){8-9}
& \normalsize{$k=1$} & \normalsize{$k=1.5$}
& \normalsize{$k=1$} & \normalsize{$k=1.5$}
& \normalsize{$k=1$} & \normalsize{$k=1.5$}
& \normalsize{$k=1$} & \normalsize{$k=1.5$} \\
\midrule
$\boldsymbol{n = 500}$  & 0.000 & 0.000 & 0.002 & 0.009 & 0.745 & 0.839 & 0.972 & 0.978 \\
$\boldsymbol{n = 1000}$ & 0.000 & 0.000 & 0.000 & 0.000 & 0.470 & 0.702 & 0.984 & 0.995 \\
$\boldsymbol{n = 2000}$ & 0.000 & 0.000 & 0.000 & 0.000 & 0.215 & 0.513 & 0.996 & 1.000 \\
$\boldsymbol{n = 4000}$ & 0.000 & 0.000 & 0.000 & 0.000 & 0.023 & 0.222 & 1.000 & 1.000 \\
$\boldsymbol{n = 8000}$ & 0.000 & 0.000 & 0.000 & 0.000 & 0.000 & 0.037 & 1.000 & 1.000 \\
\bottomrule
\end{tabular}
\caption{\footnotesize{The proportions of simulations in which the Wellek-Welch test successfully identifies the unobserved effect modifier.}}
\label{tab2}
\end{table}

\section{Full Listing of Collaborators}

Department of Biostatistics, Johns Hopkins Bloomberg School of Public Health, Baltimore, Maryland, United States (Bolun Liu, Elizabeth A. Stuart, Elizabeth L. Ogburn); Department of Mental Health, Johns Hopkins Bloomberg School of Public Health, Baltimore, Maryland, United States (Trang Quynh Nguyen); Department of Epidemiology, Johns Hopkins Bloomberg School of Public Health, Baltimore, Maryland, United States (Bryan Lau, Amii M. Kress); Department of Biostatistics, University of Michigan School of Public Health, Ann Arbor, Michigan, United States (Michael R. Elliott); Department of Psychiatry and Behavioral Sciences, University of California, San Francisco, San Francisco, California, United States (Kaja Z. LeWinn); Department of Pediatrics, School of Medicine and Public Health, University of Wisconsin–Madison, Madison, Wisconsin, United States (James E. Gern); Division of Epidemiology and Community Health, University of Minnesota, Minneapolis, Minnesota, United States (Ruby H. N. Nguyen); Department of Population Medicine, Harvard Pilgrim Health Care Institute, Boston, Massachusetts, United States (Emily Oken); Avera Research Institute, Avera Health, Sioux Falls, South Dakota, United States (Amy J. Elliott); Department of Public Health, Icahn School of Medicine at Mount Sinai, New York, New York, United States (Rosalind J. Wright); Department of Epidemiology, Dartmouth College, Hanover, New Hampshire, United States (Janet L. Peacock); Department of Statistics, University of California, Davis, Davis, California, United States (Hans\mbox{-}Georg Müller, Yidong Zhou); Department of Psychiatry, Psychology, and Clinical and Translational Science, University of Pittsburgh, Pittsburgh, Pennsylvania, United States (Alison E. Hipwell); Department of Biostatistics, Epidemiology, and Informatics, University of Pennsylvania, Philadelphia, Pennsylvania, United States (Kyle R. Busse, Ellen C. Caniglia, Yajnaseni Chakraborti, Sunni Mumford, Enrique F. Schisterman); Department of Pediatrics and Department of Environmental and Occupational Health Sciences, University of Washington, Seattle, Washington, United States (Catherine J. Karr); Feinstein Institutes for Medical Research, Northwell Health, New York, New York, United States (Arjun Sondhi); Department of Internal Medicine and Comprehensive Cancer Center, University of New Mexico, Albuquerque, New Mexico, United States (Li Luo).

\section*{Acknowledges}
\textbf{ECHO Acknowledgments:} The authors wish to thank our ECHO Colleagues; the medical, nursing, and program staff; and the children and families participating in the ECHO cohort.\\~\\
\textbf{Funding Statement and Disclaimer:} The content is solely the responsibility of the authors and does not necessarily represent the official views of the National Institutes of Health.\\~\\
Research reported in this publication was supported by the Environmental influences on Child Health Outcomes (ECHO) Program, Office of the Director, National Institutes of Health, under Award Numbers U2COD023375 (Coordinating Center), U24OD023382 (Data Analysis Center), U24OD023319 with co-funding from the Office of Behavioral and Social Science Research (Measurement Core), U24OD035523 (Lab Core), ES0266542 (HHEAR), U24ES026539 (HHEAR Barbara O’Brien), U2CES026533 (HHEAR Lisa Peterson), U2CES0\allowbreak26542 (HHEAR Patrick Parsons, Kannan Kurunthacalam), U2CES030859 (HHEAR Manish Arora), U2CES030857 (HHEAR Timothy R. Fennell, Susan J. Sumner, Xiuxia Du), U2CES026555 (HHEAR Susan L. Teitelbaum), U2CES026561 (HHEAR Robert O. Wright), U2CES030851 (HHEAR Heather M. Stapleton, P. Lee Ferguson), UG3/UH3OD023251 (Akram Alshawabkeh), UH3OD023320 and UG3OD035546 (Judy Aschner), UH3OD023332 (Clancy Blair, Leonardo Trasande), UG3/UH3OD023253 (Carlos Camargo), UG3/UH3OD0\allowbreak23248 and UG3OD035526 (Dana Dabelea), UG3/UH3OD023313 (Daphne Koinis Mitchell), UH3OD023328 (Cristiane Duarte), UH3OD023318 (Anne Dunlop), UG3/UH3OD023279 (Amy Elliott), UG3/UH3OD023289 (Assiamira Ferrara), UG3/UH3OD023282 (James Gern), UH3OD023287 (Carrie Breton), UG3/UH3OD023365 (Irva Hertz-Picciotto), UG3/UH3OD0\allowbreak23244 (Alison Hipwell), UG3/UH3OD023275 (Margaret Karagas), UH3OD023271 and UG3O\allowbreak D035528 (Catherine Karr), UH3OD023347 (Barry Lester), UG3/UH3OD023389 (Leslie Leve), UG3/UH3OD023344 (Debra MacKenzie), UH3OD023268 (Scott Weiss), UG3/UH3OD\allowbreak023288 (Cynthia McEvoy), UG3/UH3OD023342 (Kristen Lyall), UG3/UH3OD023349 (Thomas O’Connor), UH3OD023286 and UG3OD035533 (Emily Oken), UG3/UH3OD023348 (Mike O’Shea), UG3/UH3OD023285 (Jean Kerver), UG3/UH3OD023290 (Julie Herbstman), UG3/ \allowbreak UH3OD023272 (Susan Schantz), UG3/UH3OD023249 (Joseph Stanford), UG3/UH3OD023305 (Leonardo Trasande), UG3/UH3OD023337 (Rosalind Wright), UG3OD035508 (Sheela Sathyanarayana), UG3OD035509 (Anne Marie Singh), UG3OD035513 and UG3OD035532 (Annemarie Stroustrup), UG3OD035516 and UG3OD035517 (Tina Hartert), UG3OD035518 (Jennifer Straughen), UG3OD035519 (Qi Zhao), UG3OD035521 (Katherine Rivera-Spoljaric), UG3OD035527 (Emily S Barrett), UG3OD035540 (Monique Marie Hedderson), UG3OD035543 (Kelly J Hunt), UG3OD035537 (Sunni L Mumford), UG3OD035529 (Hong-Ngoc Nguyen), UG3OD035542 (Hudson Santos), UG3OD035550 (Rebecca Schmidt), UG3OD035536 (Jonathan Slaughter), UG3OD035544 (Kristina Whitworth).\\~\\
\textbf{Role of Funder Statement:} The sponsor, NIH, participated in the overall design and implementation of the ECHO Program, which was funded as a cooperative agreement between NIH and grant awardees. The sponsor approved the Steering Committee-developed ECHO protocol and its amendments including COVID-19 measures. The sponsor had no access to the central database, which was housed at the ECHO Data Analysis Center. Data management and site monitoring were performed by the ECHO Data Analysis Center and Coordinating Center. All analyses for scientific publication were performed by the study statistician, independently of the sponsor. The lead author wrote all drafts of the manuscript and made revisions based on co-authors and the ECHO Publication Committee (a subcommittee of the ECHO Operations Committee) feedback without input from the sponsor. The study sponsor did not review or approve the manuscript for submission to the journal. \\~\\
\textbf{Data Availability Statement:} Select de-identified data from the ECHO Program are available through NICHD’s Data and Specimen Hub (DASH). Information on study data not available on DASH, such as some Indigenous datasets, can be found on the ECHO study DASH webpage.
We wish to thank our ECHO colleagues; the medical, nursing, and program staff; and the children and families 

\label{lastpage}
\end{document}